\documentclass{emulateapj}

\usepackage{graphicx}
\usepackage[hypertex]{hyperref}

\def \arcmin {\hbox{$^\prime$}} 
\def \arcsec {\hbox{$^{\prime\prime}$}} 
 
\def\spose#1{\hbox to 0pt{#1\hss}} 
\def\ltsim{$\mathrel{\spose{\lower 3pt\hbox{$\sim$}} 
        \raise 2.0pt\hbox{$<$}}$\thinspace} 
\def\gtsim{$\mathrel{\spose{\lower 3pt\hbox{$\sim$}} 
        \raise 2.0pt\hbox{$>$}}$\thinspace} 
\def \msun {${\rm M_\odot}$}

\def \eg {e.g.}

\def \ie {i.e.}

\newcommand{\figc}{$f_i^{GC}$}
\newcommand{\lmxbpaper}{Paper~I}

\newcommand{\rh}{${\rm R_{1/2}}$}
\newcommand{\iraf}{{IRAF}}
\newcommand{\stsdas}{{STSDAS}}
\newcommand{\tinytim}{{TinyTim}}

\newcommand{\Sn}{${\rm S_N}$}
\newcommand{\Sl}{${\rm S_L}$}

\newcommand{\mturn}{${\rm m_V^T}$}

\newcommand{\Mturn}{${\rm M_V^T}$}

\newcommand{\chandra }{{\em Chandra}} 
\newcommand{\hst}{{\em HST}}

\newcommand{\heasoft }{{\em Heasoft}}

\newcommand{\gcinc}{${\rm f_i^{GC}}$}

\newcommand{\lgc}{${\rm L_{GC}}$}

\newcommand{\lk }{${\rm L_K}$}

\newcommand{\twomass}{{\em 2MASS}}

\newcommand{\lsun }{${\rm L_\odot}$}
 
\newcommand{\ned}{{\em{NED}}} 

\newcommand{\thin}{\thinspace}
\slugcomment{Accepted for publication in The Astrophysical Journal}
\shorttitle{Globular Clusters in Early-Type galaxies}
\shortauthors{Humphrey}

\begin{document} 
 
\title{Low-Mass X-ray Binaries and Globular Clusters in Early-Type Galaxies. II. Globular Cluster Candidates and their Mass-Metallicity Relation}
\author {{Philip J. Humphrey}\altaffilmark{1}}
\altaffiltext{1}{Department of Physics and Astronomy, University of California, Irvine, 4129 Frederick Reines Hall, Irvine, CA 92697-4575}
\begin{abstract}
We present an astrometry and photometry catalogue of globular cluster (GC) 
candidates detected with HST WFPC2 in a sample of 19 early-type galaxies,
appropriate for comparison to the low-mass X-ray binary (LMXB) populations observed
with \chandra. In a companion paper, we present the \chandra\ data and investigate
the relation between these populations. 
We demonstrate that, although there is little evidence of 
a colour-magnitude correlation for the GCs, after estimating mass and 
metallicity from the photometry, 
under the  assumption of a single age simple stellar population,
there is a significant positive correlation between mass and metallicity.
We constrained ${\rm [Z/H] = (-2.1\pm0.2)+(0.25\pm0.04)log_{10}M}$,
with a 1-$\sigma$ intrinsic scatter of 0.62~dex in metallicity.
If GCs are bimodal in metallicity this relation is consistent with recent suggestions
of a mass-metallicity relation only for metal-poor clusters.
Adopting a new technique to fit the GC luminosity function (GCLF) 
accounting for incompleteness and the Eddington bias,
we compute the V-band local GC specific frequency (\Sn) and specific 
luminosity (\Sl) of each galaxy. We show that \Sl\ is the more robust measure 
of the richness of a GC population where a significant 
fraction is undetected due to source detection incompleteness.
We find that the absolute magnitude of the GCLF turnover
exhibits intrinsic scatter from galaxy to galaxy of $\sim$0.3~mag (1-$\sigma$),
limiting its accuracy as a standard distance measure.
\end{abstract}


\section{Introduction}
Globular clusters (GCs) are found in galaxies of all morphological types
and sizes. As primarily old stellar systems,
their distribution and properties provide crucial insights into the way
in which structure forms within the Universe.
For a recent review we refer the reader to \citet{brodie06a}.
An intriguing characteristic of these objects, which provides valuable clues
as to their structure and internal dynamics, is their association with 
low-mass X-ray binaries (LMXBs). Although it has long been recognized that 
LMXBs are overabundant per unit optical light by a factor $\sim$100 in Milky
Way GCs as compared to the field \citep{fabian75a,clark75a}, the small number
of sources has limited what this phenomenon can tell us about LMXB formation.
With the advent of \chandra, however, it has become feasible to resolve 
individual LMXBs in galaxies outside the Local Group, and thus to begin to 
assemble larger samples of LMXB-hosting GCs \citep[\eg][]{sarazin03}. 
Given their typically rich GC populations 
and their clean, old stellar populations which prevent contamination of the 
X-ray sources with high-mass X-ray binaries, massive 
early-type galaxies, in particular, provide an ideal environment 
in which to conduct such studies 
\citep[\eg][]{kundu02a,sarazin03,humphrey04a}.

Even with the small samples of galaxies in which the GC-LMXB connection
has been investigated to date, some intriguing trends have already been observed.
By comparing the numbers of GCs and the total luminosity of LMXBs 
in early-type galaxies \citet{irwin05a} argued that a significant fraction
of the LMXBs observed in the field form in the field, despite the stellar
population being old. \citet{juett05a} reported a similar result although,
since they did not strictly compare the numbers of LMXBs and GCs within the same
apertures, the strength of their correlation has been called into
question \citep[][]{kim05a}.
Typically $\sim$4\% of GCs are observed to harbour LMXBs, with brighter
and redder (more metal rich) GCs preferentially likely to contain them
\citep{kundu02a,sarazin03,kim05a}.
The GC luminosity dependence is broadly consistent with the probability 
of harbouring an LMXB being proportional to the mass or to the stellar 
capture cross-section in the GC \citep[\eg][]{jordan04a,smits06a,sivakoff06a}.
The origin of the colour dependence is unclear and several different possible
explanations have been proposed \citep[for a review of some of these, see][]{jordan04a}.
\citet{maccarone04a} suggested it may relate
to different patterns of mass-transfer in metal-rich and metal-poor
LMXBs. Alternatively, it may arise due to variations in the IMF between 
metal-poor and metal-rich clusters \citep{grindlay87a}, or the efficiency of 
magnetic braking \citep{ivanova06c}.
So far, however, most studies of the LMXB-GC connection
have been relatively small and so strong general conclusions are difficult 
to draw.

The properties of the GC populations in early-type galaxies 
themselves are also of considerable interest since they provide a unique insight 
into how such galaxies form.
Although the numbers of GCs do not simply scale with the total stellar mass
of a galaxy, the GC luminosity function (GCLF) is observed to be remarkably
uniform \citep[\eg][]{harris91a}. Typically it can be well-fitted by a log-normal
distribution, the absolute shape of which varies only weakly with
the galaxy properties \citep[\eg][]{kundu01b,jordan06a}.
The origin of this shape may be related to the dynamical destruction 
of low-mass GCs \citep[\eg][]{vesperini03a}.
The stability of the GCLF peak (``turnover'') has led to its adoption 
as a standard distance measure for nearby galaxies \citep{harris91a,jacoby92},
although there has been some debate as to its reliability
\citep[\eg][]{ferrarese00a,kundu01b}. 
A recent study of early-type galaxies in Virgo (which span 
$\sim$7~magnitudes in B) by \citet{jordan07a} found that the width and,
possibly, the turnover magnitude vary with the galaxy absolute magnitude,
in the sense that the least-massive galaxies have narrower, fainter GCLFs.
The authors did, however, find significant scatter about these relations
for galaxies at a given magnitude. 

The GC colour distributions of early-type galaxies 
are observed to be almost universally bimodal,
\citep[\eg][]{zepf93a,gebhardt99a}, probably indicating ubiquitous metal-rich and 
metal-poor sub-populations \citep[although see][]{richtler06a,yoon06a}.
The GC distribution, in particular that of the metal-poor (\ie\ blue) 
population, is observed to be substantially more extended than the optical 
light \citep{geisler96a}.
The average colour of the GCs in a galaxy strongly correlates
with the total galaxy magnitude \citep{brodie91a}, 
reflecting an increasing fraction of red GCs in more massive galaxies
coupled with a correlation between mean GC metallicity and galaxy mass for both blue 
and red clusters \citep{peng06a}.
There is, however, no correspondingly strong 
correlation between the colour and luminosity of the GCs themselves 
\citep[\eg][]{larsen01a,kundu01b}.
A weak colour-magnitude relation has been observed, but only
in  metal-poor GCs, in recent observations of nearby galaxies 
\citep[][]{harris06a,strader06a,mieske06a}.
These clues point to different origins for the two sub-populations, with the
metal-poor clusters possibly forming in low-mass halos in the very early Universe
and being accreted during hierarchical structure formation. In contrast,
metal rich clusters may form during the gas-rich mergers which assemble the 
massive galaxy \citep[][and references therein]{brodie06a}.

In this paper, we present a study of the GC populations in 19 early-type galaxies
observed with \hst\ WFPC2. In particular, we present a photometric catalogue 
of GC candidates 
which can be directly compared to the properties of observed LMXBs.
In a companion paper, \citet[][hereafter \lmxbpaper]{humphrey06b}, we present
the \chandra\ data for these galaxies and investigate the relation between
these populations, significantly expanding the sample of galaxies in which
the LMXB-GC connection has been investigated.
In the present work we compute self-consistently derived properties of
the GC populations, such as the GC specific frequency, \Sn, 
specific luminosity, \Sl, and the GCLF turnover, \mturn.
The self-consistent computation of such parameters is essential for our
analysis in Paper~I.
The galaxies used in this study were chosen from the \chandra\ and \hst\ 
archives as having sufficiently deep ACIS and  WFPC2 
data to enable a significant fraction of the LMXBs and GCs to be individually
resolved. 
To enable a consistent comparison between the galaxies and to 
mitigate against the rapidly-degrading \chandra\ PSF off-axis, we focus here
only on WFPC2 pointings of the centre of each galaxy.
The galaxies, and the details of the archival observations used are
shown in Table~\ref{table_obs}.
All errors quoted are 90\% confidence regions, unless
otherwise stated in the text.

\section{Observations and data analysis} \label{sect_reduction}
\renewcommand{\tabcolsep}{1.0mm}
\begin{deluxetable*}{llllllllllllll}
\tabletypesize{\scriptsize}
\tablecaption{Target list and observation log \label{table_obs}}
\tablehead{
\colhead{Name}& \colhead{Type} & \colhead{Dist} & \colhead{\lk} & 
\multicolumn{3}{l}{B} & \multicolumn{3}{l}{V} & \multicolumn{3}{l}{I} 
& \colhead{R$_{min}$}\\
\colhead{} & \colhead{} & \colhead{} & \colhead{} & 
\colhead{Filt} & \colhead{ObsID} & \colhead{Exp} &
\colhead{Filt} & \colhead{ObsID} & \colhead{Exp} &
\colhead{Filt} & \colhead{ObsID} & \colhead{Exp} & \colhead{(\arcsec)}\\
\colhead{} & \colhead{} & \colhead{(Mpc)} & \colhead{($10^{11}$\lsun)} & 
\colhead{} & \colhead{} & \colhead{(s)} &
\colhead{} & \colhead{} & \colhead{(s)} &
\colhead{} & \colhead{} & \colhead{(s)} 
}
\startdata
NGC1332 & S(s)0 & $21.3$  & $1.4$  & \ldots & \ldots & \ldots & F555W & u2tv03 & 160 & F814W & u2tv03 & 320 & $2.9$  \\
NGC1387 & SAB(s)0 & $18.9$  & $0.78$  & \ldots & \ldots & \ldots & F606W & u29r5h & 160 & \ldots & \ldots & \ldots & $8.7$  \\
NGC1399 & cD & $18.5$  & $2.1$  & F450W & u34m02 & 2600 & \ldots & \ldots & \ldots & F814W & u34m02 & 1200 & $0.31$  \\
NGC1404 & E1 & $19.5$  & $1.5$  & F450W & u34m04 & 2400 & \ldots & \ldots & \ldots & F814W & u34m04 & 1260 & $1.7$  \\
NGC1553 & SA(rl)0 & $17.2$  & $1.9$  & \ldots & \ldots & \ldots & F555W & u2tv28 & 160 & F814W & u2tv28 & 320 & $2.0$  \\
NGC3115 & S0 & $9.00$  & $0.74$  & \ldots & \ldots & \ldots & F555W & u2j20b & 860 & F814W & u2j20b & 1236 & $2.7$  \\
NGC3585 & E7/S0 & $18.6$  & $1.5$  & \ldots & \ldots & \ldots & F555W & u3m719 & 800 & F814W & u3m719 & 1800 & $3.7$  \\
NGC3607 & SA(s)0 & $21.2$  & $1.5$  & \ldots & \ldots & \ldots & F555W & u2tv12 & 160 & F814W & u2tv12 & 320 & $8.4$  \\
NGC4125 & E6pec & $22.2$  & $1.8$  & \ldots & \ldots & \ldots & F555W & u3m723 & 1400 & F814W & u3m723 & 2100 & $7.4$  \\
NGC4261 & E2-3 & $29.3$  & $2.2$  & \ldots & \ldots & \ldots & F547M & u2i502 & 800 & F791W & u2i502 & 800 & $1.6$  \\
NGC4365 & E3 & $19.0$  & $1.6$  & \ldots & \ldots & \ldots & F555W & u2bm07 & 1000 & F814W & u2bm07 & 460 & $3.5$  \\
NGC4472 & E2/S0(2) & $15.1$  & $3.2$  & \ldots & \ldots & \ldots & F555W & u66306 & 7200 & F814W & u66315 & 5200 & $2.9$  \\
NGC4494 & E1-2 & $15.8$  & $0.81$  & \ldots & \ldots & \ldots & F555W & u3vz05 & 1500 & F814W & u3vz05 & 1800 & $2.3$  \\
NGC4552 & E & $14.3$  & $0.85$  & \ldots & \ldots & \ldots & F555W & u30708 & 1800 & F814W & u30708 & 1500 & $0.85$  \\
NGC4621 & E5 & $17.0$  & $1.2$  & \ldots & \ldots & \ldots & F555W & u2j20d & 1380 & F814W & u2j20d & 830 & $6.4$  \\
NGC4649 & E2 & $15.6$  & $2.5$  & \ldots & \ldots & \ldots & F555W & u2qo03 & 2100 & F814W & u2qo03 & 2500 & $87.$  \\
NGC5018 & E3 & $42.6$  & $3.0$  & \ldots & \ldots & \ldots & F555W & u3m725 & 800 & F814W & u3m725 & 1800 & $4.3$  \\
NGC5845 & E & $24.0$  & $0.27$  & \ldots & \ldots & \ldots & F555W & u30709 & 2140 & F814W & u30709 & 1120 & $3.8$  \\
NGC5846 & E0-1 & $21.1$  & $1.5$  & \ldots & \ldots & \ldots & F555W & u36j04 & 1300 & F814W & u36j04 & 1400 & $5.4$ 
\enddata
\tablecomments{Summary of the sample galaxies and the observations used in the
present analysis. The morphological type (Type) is taken from \ned.
 All distances (Dist) are  taken
from  \citet{tonry01}, corrected by -0.16~mag to account for revisions to the Cepheid
zero-point \citep[\eg][]{jensen03}, except that of NGC\thin 5018, which was taken from
\citet{faber89}. Total, extrapolated 
K$_s$-band luminosities (\lk) were taken from \twomass\ \citep{jarrett00a}, adopting $M_{K\odot}=3.41$. The zero-point was chosen for consistency with the
stellar population models of \citet{maraston05a} (\S~\ref{sect_mass_metallicity}) 
and assuming the 
K-band and K$_s$ band magnitudes do not differ significantly \citep[\eg][]{carpenter01a}.
 For each standard UBV filter band, we
list the observation ID (ObsId) of the WFPC2 datasets used, the actual filter used
(Filt) and the total exposure in seconds (Exp). In addition, we show the approximate
semi-major axis of the central, excluded region (R$_{min}$).}
\end{deluxetable*}
Data reduction and analysis were performed using the 
{\em Image Reduction and Analysis Facility}, \iraf\ (PC-IRAF vers. 2.12.2a), and
the {\em Space Telescope Science Data Analysis System}, \stsdas\ (vers. 3.2)
software suites. The individual pipelined images were first processed to correct for
warm pixels with the {\em warmpix} \stsdas\ task. 
The corrected images were then aligned and, where the gain and filters were identical,
combined with the \stsdas\ task {\em\ gcombine}
to remove cosmic ray events. The alignment was achieved initially by examining the 
astrometry keywords in the data-file and shifting individual images an appropriate integer
number of pixels. In a few cases the astrometry keywords were insufficiently accurate
and so the offset between the images was determined by matching moderately bright
point-sources between the exposures.
In order to remove residual cosmic ray events and defective pixels which have not been
removed in earlier processing (in particular since there are some pointings for which only
a single exposure is available), the resulting images were processed with the 
LA-Cosmic\footnote{Available at {http://www.astro.yale.edu/dokkum/lacosmic}} algorithm of \citet{vandokkum01a}.

The detection of point sources against the bright, strongly-varying, ``background'' of 
galaxy emission is particularly challenging. To achieve this, we adopted a procedure
similar to that outlined by \citet[][\citealt{kundu01b}]{kundu99a}. First, to improve S/N, for each 
WFPC2 
CCD we aligned and combined the images accumulated with different filters. An initial
source list was created with the \iraf\ task {\em daofind}, having first estimated
the statistical noise level from the region of the CCD with the least background 
galaxy emission and free from obvious point-sources. Due to the rapidly rising background
across the WFPC2 field of view this procedure yields a very large number of detections,
most of which are false on each CCD. To eliminate false detections, we recomputed the 
S/N for each source, based on local background estimates. To measure the source
counts, we adopted a 3 pixel radius aperture and we estimated 
the background from an annulus with inner and outer radii of 3 and 5 pixels. 
We rejected any source with S/N lower than 4.8, implying only $\sim$1 false detection
per CCD. In order to provide useful colour information, we required that any source
was detected in all filters used. 
To eliminate spatially extended sources (most likely background galaxies), 
we excluded any source with a ``concentration'' (defined as
the ratio of background-subtracted counts within a 3 pixel aperture to that within 
a 0.5 pixel aperture) outside the range 2--10. 
All detections were confirmed by visual inspection of the images.

To minimize confusion with small-scale but extended features (such as central
dust-lanes) in the galaxies, which seriously complicate the computation of
background-subtracted photometry, 
we eliminated detections within the central few arcseconds of each galaxy (shown
in Table~\ref{table_obs}). In particular, data in the vicinity of a central 
nuclear disc 
in NGC\thin 1387 and the bright knots of emission in the centre of NGC\thin 4125
were excluded. The projection of the bright spiral NGC\thin 4647 partly overlaps
the image of NGC\thin 4649, and so to minimize confusion
we excluded any sources lying within the
B-band twenty-fifth magnitude ellipse of NGC\thin 4647,
as listed in \citet{devaucouleurs91}.

Photometry was obtained with a 3 pixel radius aperture for the PC, and a 
2 pixel aperture for the WF CCDs. The background was accumulated from an
annulus between 3 and 5 pixels in radius. 
Since a GC may be slightly resolved in the case of the galaxies here, the proximity of 
the background annulus to the centre of the GC may lead to a slight over-subtraction
and hence underestimate of the true brightness of a large GC. To estimate the likely
magnitude of this effect, we simulated images of a 
GC at 15~Mpc, adopting a \citet{king62a} model for its surface brightness profile. We 
adopted a ``concentration'' (the logarithm of the ratio between the truncation and
core radii) of 1.25 and a half-light radius of 2.5~pc \citep[\eg][]{kundu99a}. The
images were convolved with images of the PSF for the wide field and planetary cameras,
generated with the  \tinytim\footnote{Available at {http://www.stsci.edu/software/tinytim/}} 
software package. For the planetary and wide field cameras, the magnitude of over-subtraction
was $\sim$0.1~mag and $\sim$0.02~mag respectively. This magnitude is similar to the 
\ltsim 0.1~mag error on the total magnitude of the GC estimated by \citet{kundu99a} on account of 
typical GCs being slightly resolved at $\sim$10--20~Mpc.
These small effects will not significantly affect the interpretation of our results.
Photometric zero-points were taken from \citet{holtzman95}, taking into account
the appropriate gain correction. Aperture correction factors were estimated 
based on simulations using  \tinytim\
for each CCD, assuming the spectrum of a K4V star, and generally agreed with the 
calibration of \citet{holtzman95b}. 
To estimate the extinction correction, we converted the line-of-sight estimates 
of \citet[][listed in \ned]{schlegel98a} to the wavelength of the mid-point 
of the appropriate filter, using the extinction law of \citet{cardelli89a}.
This was typically very close to the \citeauthor{schlegel98a} value for the nearest 
standard photometric band. 
The photometry was converted to the standard UBVRI system following the conversion
formulae given in \citet{holtzman95}, assuming a standard V-I colour of 1.14 and 
a B-V colour of 0.97.
For a consistent comparison between the galaxies, we converted the (B,I) band photometry 
of NGC\thin 1399 and NGC\thin 1404 into (V,I) band photometry by adopting the conversion 
formula given in \citet{gebhardt99a}.

\section{GC Colours}
\subsection{Colour Histograms}
\begin{figure}
\centering
\includegraphics[scale=0.35,angle=270]{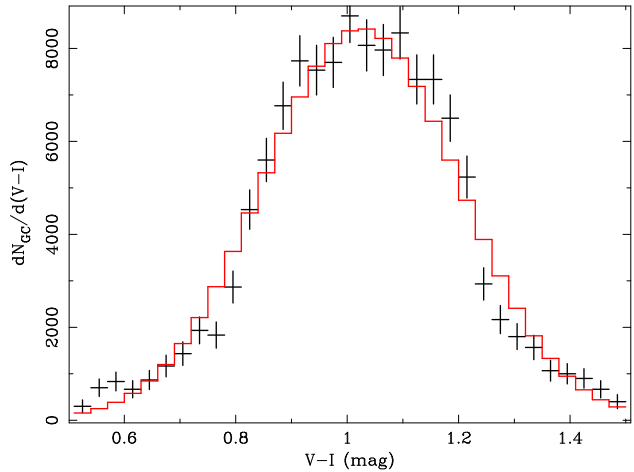}
\caption{Composite GC colour distribution, fitted with a single Gaussian
model. Although the model does not capture all of the fine detail of the 
distribution (which may be bimodal), 
it is a reasonable approximation of the overall shape.} 
\label{fig_colour_hist}
\end{figure}
In order to investigate the colour distribution of GCs,
we computed histograms of V-I colour for each galaxy. In \lmxbpaper\
we investigate the dependence of the probability that a GC hosts an LMXB as 
a function of metallicity, which we estimate by a simple transformation of 
the GC colour. Although, in general the colour distributions of GCs in early-type 
galaxies are bimodal, we do not expect whether a GC belongs to the red or 
blue subpopulation to affect its likelihood to host an LMXB, but rather this should
depend on the metallicity. To obtain a global fit to this metallicity dependence
in \lmxbpaper, we therefore need only to parameterize the composite 
colour distribution of all the GCs
with a model which describes its {\em overall} shape reasonably well,
and we do not need to capture the fine detail. As a simple approximation,
we chose to adopt a single Gaussian model, which has previously been fitted to
such data \citep[\eg][]{gebhardt99a}. We show in Fig~\ref{fig_colour_hist}
the Gaussian fit to the composite GC colour histogram for all of the galaxies,
with a mean V-I of 1.033$\pm0.005$ and $\sigma$ of 0.186$\pm0.005$. Although
the fit is formally unacceptable ($\chi^2$/dof=57/27), as is clear from 
Fig~\ref{fig_colour_hist} the model does describe the global shape of the 
distribution reasonably well; in fact it is accurate to
$\sim$15\% for most of the interesting range of colours. This  is adequate
for our purposes in \lmxbpaper, as evinced by the excellent agreement we 
find in that paper between our measured metallicity dependence and that
found by other authors with different techniques.

For comparison with other authors' work, and to facilitate the comparison
between galaxies  in \lmxbpaper, we also fitted the single Gaussian model
to the GC colour distributions of each galaxy. Given the fewer numbers of 
sources in each bin, we adopted a Cash-C minimization algorithm to fit the model
to the data, and assessed the goodness-of-fit {\em via} 100 Monte-Carlo simulations.
For each simulation, an artificial dataset was created given the best-fitting number
of sources in each colour bin, which was then fitted with our preferred model. The 
fraction of simulations yielding a higher  Cash-C statistic (\ie\ worse fit) than
the model fit to the real data can be used as an estimate of the goodness of fit.
The fit was formally acceptable at a 5\% significance level in most of the galaxies.
Unacceptable fits were found for NGC\thin 1399, NGC\thin 3115, NGC\thin 4494, 
NGC\thin 4552, NGC\thin 4621 and NGC\thin 4649, which typically reflects evidence of 
bimodality or asymmetry in the GC colour distribution.
Nonetheless, the fits of the single Gaussian model to the GC colour distributions 
of each galaxy (shown in Table~\ref{table_derived}) generally
yield results consistent with those of
\citet{gebhardt99a}, \citet{kundu01b,kundu01} or \citet{larsen01a},
where our sample overlapped with theirs.

\subsection{Mass-metallicity relation} \label{sect_mass_metallicity}
\begin{figure*}
\centering
\includegraphics[scale=0.38]{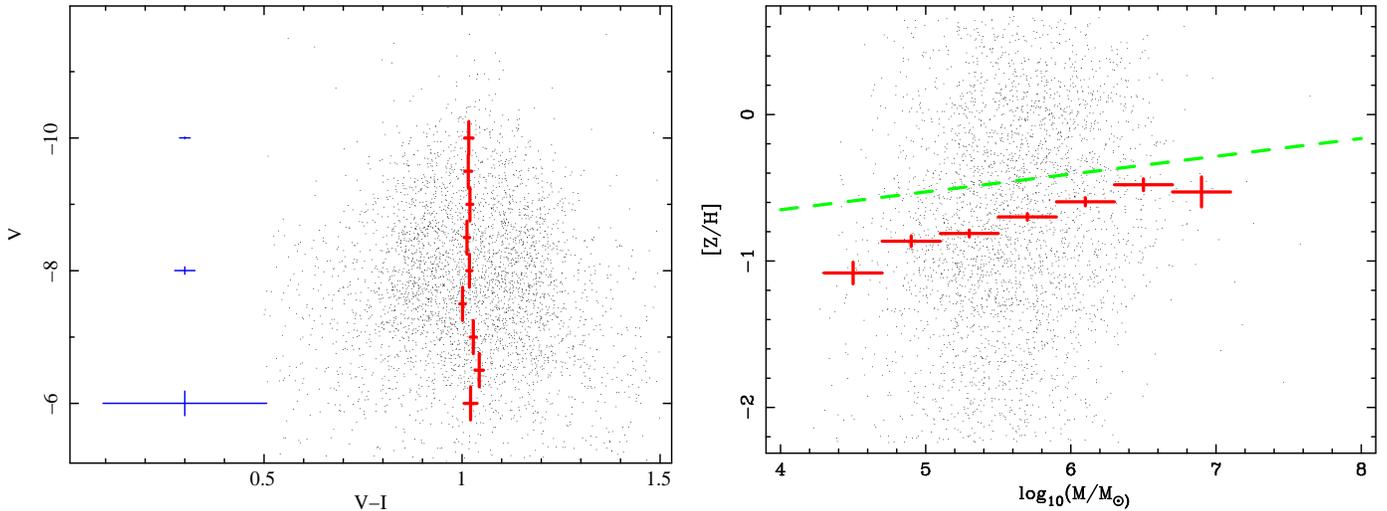}
\caption{Left: Composite colour-magnitude diagram for all of the GC candidates.
We show (blue) representative error-bars for a range of V magnitudes.
Overlaid we show the data-points binned into a series of magnitude bins,
clearly indicating little or no colour-magnitude relation. 
 Right: 
the same data, transformed into mass and metallicity space (see text).
Overlaid we show the same data binned into a series of metallicity bins,
clearly indicating a slight mass-metallicity correlation. We also show (dotted line)
an extrapolation of the best-fitting mass-metallicity relation for early-type galaxies 
found by
\citet{thomas05a}} 
\label{fig_composite_cmd}
\end{figure*}
We show in Fig~\ref{fig_composite_cmd} (left panel) the composite 
colour-magnitude diagram
for all of the GCs detected in more than one filter. 
We also show representative error-bars and the data rebinned into a series
of narrow magnitude bins. The data do
not show convincing evidence of a significant colour-magnitude correlation.
For example, the binned-up data
are consistent ($\chi^2$/dof=$13.6/8$) with no variation in
colour with magnitude. The fit does not improve significantly ($\chi^2$=$12.2/7$)
when fitted with a linear model 
of the form $(V-I) = (1.043^{+0.014}_{-0.005}) + (0.003\pm0.005) V$; this is 
consistent with no variation within errors and indicates that, if any correlation
is present, it is extremely weak.
The V-I colour used in the present work does not cleanly separate the metal-poor
and metal-rich sub-populations of GCs;  coupled with the small field of view of the 
WFPC2, which limits the numbers of blue GCs detected, we do not see
the so-called ``blue tilt'' (\ie\ colour-magnitude correlation) recently observed in the 
metal-poor GC populations of several galaxies  \citep{harris06a,strader06a}.

In the right panel of Fig~\ref{fig_composite_cmd}, we show the same data converted into mass and metallicity. 
To achieve this, we linearly interpolated the V and I-band photometry onto the 
updated simple stellar population (SSP) model grids of 
\citet[][see also \citealt{maraston98a}]{maraston05a}
\footnote{http://www-astro.physics.ox.ac.uk/$\sim$maraston/Claudia's\_Stellar \_Population\_Models.html}. We adopted these models for consistency with our previous study of 
the mass and metallicity of early-type galaxies, in which we adopted the alpha-enhanced
Lick index models of \citet{thomas03a}, built upon these SSP models
\citep{humphrey05a}. In our study of the mass profiles of early-type galaxies, 
\citet{humphrey06a}, we 
found that our measured K-band mass-to-light ratios in the (baryon-dominated)
centres of the galaxies were consistent with the predictions of the 
\citeauthor{maraston05a} models, provided one adopted a \citet{kroupa01a} initial mass function,
which we therefore adopted in the present work.
\citeauthor{maraston05a} provides the mass-to-light ratio as a function of 
metallicity and age. We assumed an age of 13~Gyr, and adopted the SSP models
computed for a \citeauthor{kroupa01a} initial mass function and a blue horizontal branch (HB) 
for the GC stars. A small fraction of the GCs with best-fitting abundances pegged at 
the high or low boundaries ([Z/H]=-2.25 and 0.67, respectively) were discounted.
Since the V and I-band mass-to-light ratios 
 show a dependence upon the metallicity, the {\em lack}
of a correlation between V-I and V actually translates into a slight mass-metallicity
correlation. Formal correlation testing was carried out using three different 
methods, Pearson's linear correlation test, Spearman's rank-order test and 
Kendal's $\tau$ test. All three tests indicated that the probability of no correlation
is less than $10^{-29}$.

We also show in Fig~\ref{fig_composite_cmd} the data
binned into a series of narrow mass bands. The error-bars on the mean
GC metallicity in each bin were estimated from the 1-$\sigma$ scatter of the 
data-points, using  the central limit theorem. 
These data are well-fitted ($\chi^2$/dof=4.3/5) by a simple linear model of the form
${\rm [Z/H] = (-2.1\pm0.2)+(0.25\pm0.04)log_{10}M}$, where M is the mean GC mass.
To estimate the intrinsic scatter about this best-fitting relation, we additionally
fitted this model to the y-on-x data-points for each GC, taking into account the errors 
on both axes with an algorithm similar to that outlined in \citet[][see also \citealt{tremaine02a}]{nr}. We added in
quadrature with the metallicity error an additional, constant, systematic error-term,
which was adjusted until the $\chi^2$/dof value was 1, 
at which point its value is approximately equal to the 
1-$\sigma$ intrinsic scatter about the best-fitting
linear relation. In this way, we estimated this scatter to be 0.62~dex in metallicity.

Since the assumption of a blue HB is most appropriate for metal-poor GCs, we have
investigated the impact of using the red HB models instead of the blue models.
We found that this produces a small shift in [Z/H], at maximum $\sim$0.2~dex,
which does not appear to correlate with mass. We found similar shifts (albeit
typically of a different sign) if we adopted a GC age of 10~Gyr. We therefore
conclude that relaxing our assumptions of a fixed GC age or a blue HB for 
all GCs is unlikely to make the observed trend disappear.
\begin{figure}
\centering
\includegraphics[scale=0.35]{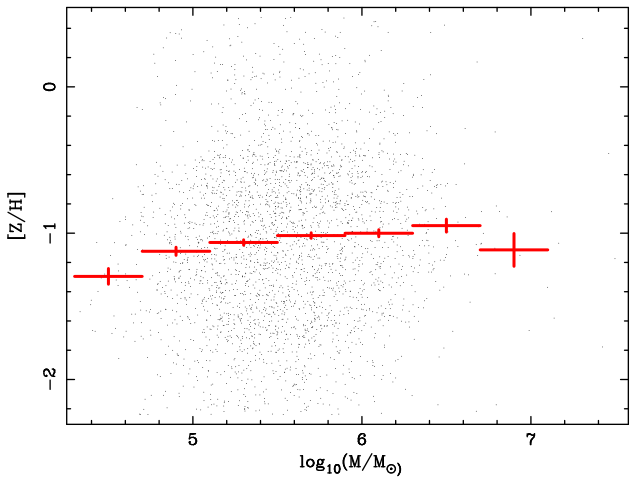}
\caption{Same as Fig~\ref{fig_composite_cmd}, right panel, but using the PEGASE
SSP models to convert colour and magnitude to mass and metallicity.} 
\label{fig_pegase_cmd}
\end{figure}
In order to verify that our results do not sensitivity depend upon the choice of 
SSP models, we also experimented with adopting models based on the 
{\bf PEGASE.2} code of \citet{fioc97a,fioc99a} to perform the transformation from
colour to mass and metallicity space. We took most of the default parameters of the code,
adopted a \citeauthor{kroupa01a} IMF, 
assumed a single burst of star-formation and we included the mass of neutron-stars and 
white dwarfs in the calculation of the mass-to-light ratio. 
For consistency with the  \citeauthor{maraston05a} models, we adopted approximately
the same transformation between the heavy element mass fraction (``Z'') 
and [Z/H] as those authors. The resulting mass-metallicity plot is shown in 
Fig~\ref{fig_pegase_cmd}. We also show the data binned into a series of narrow mass bands, which clearly
indicates a mass-metallicity correlation similar to that found above. 
Fitting the binned data with a simple linear model, we found
that the data are inconsistent with a constant [Z/H] model ($\chi^2$/dof=42/6),
but are in agreement ($\chi^2$/dof=11/5) with a model of the form 
${\rm [Z/H] = (-1.8\pm0.2)+(0.12\pm0.03)log_{10}M}$.
Both the slope and normalization of the best-fitting model
differ from the relation obtained with the \citeauthor{maraston05a} models, which reflects
quantitative differences in the relation between the mass-to-light ratios and 
the metallicity for the two sets of models.

\section{Luminosity Functions}
\begin{deluxetable*}{llllllllll}
\tablecaption{Derived Properties of the GC Populations \label{table_derived}}
\tabletypesize{\scriptsize}
\tablehead{
\colhead{Name }&\colhead{$N_{GC}^{det}$ }&\colhead{$V-I_0$ }&\colhead{$\sigma_{V-I}$ }&\colhead{\gcinc }&\colhead{\lgc }&\colhead{$L_V^{FOV}$ }&\colhead{\mturn }&\colhead{\Sn }&\colhead{\Sl } \\
\colhead{ }&\colhead{ }&\colhead{ }&\colhead{ }&\colhead{ }&\colhead{($10^7$\lsun)}&\colhead{($10^{10}$\lsun)}&\colhead{ }&\colhead{ }&\colhead{ }}
\startdata
NGC1332 &$198$ &$0.998\pm 0.03$ &$0.22^{+0.03}_{-0.02}$ &$0.60$ &$8.3\pm 1.2$ &$2.0\pm 0.9$ &$24.4\pm 0.4$ &$2.5\pm 1.1$ &$0.42\pm 0.19$ \\
NGC1387 &$28$ &\ldots &\ldots &$0.22$ &$0.79\pm 0.51$ &$1.2\pm 0.3$ &\ldots &$0.38\pm 0.26$ &$0.065\pm 0.045$ \\
NGC1399 &$467$ &$1.034\pm 0.009$ &$0.123^{+0.007}_{-0.006}$ &$0.81$ &$12.\pm 1.$ &$2.0\pm 1.8$ &$24.0\pm 0.2$ &$3.5\pm 3.1$ &$0.59\pm 0.53$ \\
NGC1404 &$109$ &$1.01\pm 0.02$ &$0.14\pm 0.01$ &$0.74$ &$5.3\pm 0.9$ &$1.7\pm 1.8$ &$24.3\pm 0.4$ &$1.8\pm 2.0$ &$0.32\pm 0.35$ \\
NGC1553 &$71$ &$0.996\pm 0.04$ &$0.19^{+0.04}_{-0.03}$ &$0.76$ &$2.1\pm 0.5$ &$2.7\pm 0.6$ &$24.5\pm 0.6$ &$0.45\pm 0.14$ &$0.077\pm 0.024$ \\
NGC3115 &$136$ &$1.01\pm 0.02$ &$0.15^{+0.02}_{-0.01}$ &$0.92$ &$2.3\pm 0.4$ &$1.1\pm 0.1$ &$22.7\pm 0.3$ &$1.2\pm 0.3$ &$0.21\pm 0.04$ \\
NGC3585 &$90$ &$0.99\pm 0.04$ &$0.23^{+0.05}_{-0.03}$ &$0.83$ &$2.4\pm 0.4$ &$3.0\pm 0.8$ &$24.6\pm 0.4$ &$0.47\pm 0.15$ &$0.081\pm 0.026$ \\
NGC3607 &$106$ &$0.95\pm 0.03$ &$0.19^{+0.03}_{-0.02}$ &$0.62$ &$4.9\pm 0.9$ &$2.4\pm 0.8$ &$24.0\pm 0.4$ &$1.2\pm 0.4$ &$0.20\pm 0.07$ \\
NGC4125 &$196$ &$1.05\pm 0.03$ &$0.24^{+0.04}_{-0.03}$ &$0.74$ &$2.6\pm 0.4$ &$3.2\pm 0.8$ &$25.5\pm 0.3$ &$1.3\pm 0.5$ &$0.082\pm 0.023$ \\
NGC4261 &$241$ &$1.00\pm 0.02$ &$0.20\pm 0.02$ &$0.56$ &$11.\pm 1.$ &$3.4\pm 3.4$ &$24.6\pm 0.3$ &$2.0\pm 2.0$ &$0.34\pm 0.34$ \\
NGC4365 &$291$ &$1.01\pm 0.02$ &$0.18\pm 0.01$ &$0.81$ &$8.2\pm 0.9$ &$2.3\pm 0.6$ &$24.0\pm 0.2$ &$2.1\pm 0.6$ &$0.36\pm 0.11$ \\
NGC4472 &$397$ &$1.08\pm 0.02$ &$0.18\pm 0.01$ &$0.83$ &$7.1\pm 0.7$ &$3.9\pm 0.5$ &$23.8\pm 0.2$ &$1.1\pm 0.2$ &$0.18\pm 0.03$ \\
NGC4494 &$155$ &$0.98\pm 0.02$ &$0.18\pm 0.02$ &$0.86$ &$3.0\pm 0.4$ &$1.5\pm 0.5$ &$23.5\pm 0.2$ &$1.2\pm 0.4$ &$0.20\pm 0.07$ \\
NGC4552 &$240$ &$1.02\pm 0.02$ &$0.18^{+0.02}_{-0.01}$ &$0.85$ &$4.2\pm 0.5$ &$1.3\pm 0.4$ &$23.4\pm 0.2$ &$1.9\pm 0.6$ &$0.32\pm 0.10$ \\
NGC4621 &$173$ &$1.04\pm 0.02$ &$0.15^{+0.02}_{-0.01}$ &$0.86$ &$3.9\pm 0.5$ &$1.7\pm 0.5$ &$23.4\pm 0.2$ &$1.3\pm 0.4$ &$0.23\pm 0.07$ \\
NGC4649 &$323$ &$1.05\pm 0.02$ &$0.17\pm 0.01$ &$0.87$ &$6.8\pm 0.7$ &$2.9\pm 0.4$ &$23.6\pm 0.2$ &$1.4\pm 0.3$ &$0.23\pm 0.04$ \\
NGC5018 &$90$ &$0.87^{+0.04}_{-0.05}$ &$0.21^{+0.05}_{-0.03}$ &$0.47$ &$7.8\pm 1.8$ &$9.7\pm 4.1$ &$25.1^{+0.6}_{-0.5}$ &$0.46\pm 0.22$ &$0.080\pm 0.038$ \\
NGC5845 &$45$ &$0.96^{+0.08}_{-0.15}$ &$0.29^{+0.19}_{-0.07}$ &$0.47$ &$1.4\pm 0.8$ &$1.9\pm 1.1$ &\ldots &$0.45\pm 0.36$ &$0.077\pm 0.062$ \\
NGC5846 &$288$ &$1.04\pm 0.02$ &$0.18^{+0.02}_{-0.01}$ &$0.76$ &$5.3\pm 0.6$ &$2.6\pm 1.1$ &$24.8\pm 0.2$ &$1.4\pm 0.6$ &$0.21\pm 0.09$ \\
\enddata
\tablecomments{Derived properties of each of the galaxies' GC distributions. 
We list the number of 
GCs detected ($N_{GC}^{det}$), the mean colour of the GCs (V-I$_0$), the 
half-width of the V-I distribution ($\sigma_{V-I}$), the total fraction
of the GC luminosity which has been detected ($f_i^{GC}$), the total V-band luminosity
of the GCs, correcting for incompleteness ($L_{GC}$), the V-band magnitude of the
galaxy in the field of view ($L_{V}^{FOV}$), 
the V-band GCLF turnover (\mturn), the specific frequency of GCs (\Sn)
and the specific luminosity (\Sl). The correlations between
these properties and the LMXB distribution are discussed in detail in \lmxbpaper.
All quoted errors are 90\%\ confidence limits, incorporating only statistical uncertainties.
An assessment of statistical errors is given in Table~\ref{table_syserr}.} 
\end{deluxetable*}

\begin{deluxetable*}{llllllllllll}
\tablecaption{Systematic error-budget for derived parameters \label{table_syserr}}
\tabletypesize{\scriptsize}
\tablehead{
\colhead{Name }&\multicolumn{2}{l}{\gcinc}& \multicolumn{3}{l}{\lgc}& \multicolumn{3}{l}{\Sn}&\multicolumn{3}{l}{\Sl} \\
\colhead{ }&\colhead{$\Delta_{free}$ }&\colhead{$\Delta_{wid}$ }&\colhead{$\Delta_{stat}$ }&\colhead{$\Delta_{free}$ }&\colhead{$\Delta_{wid}$ }&\colhead{$\Delta_{stat}$ }&\colhead{$\Delta_{free}$ }&\colhead{$\Delta_{wid}$ }&\colhead{$\Delta_{stat}$ }&\colhead{$\Delta_{free}$ }&\colhead{$\Delta_{wid}$ } \\
}
\startdata
NGC1332 &$-0.052$ &$-0.063$ &$1.2$ &$0.83$ &$0.98$ &$1.1$ &$1.5$ &$1.4$ &$0.19$ &$0.042$ &$0.050$ \\
NGC1387 &\ldots &$-4.8\times 10^{-4}$ &$0.51$ &\ldots &$3.9\times 10^{-3}$ &$0.26$ &\ldots &$0.71$ &$0.045$ &\ldots &$3.2\times 10^{-4}$ \\
NGC1399 &$-0.021$ &$-0.049$ &$1.0$ &$-0.50$ &$0.88$ &$3.1$ &$0.97$ &$1.0$ &$0.53$ &$-0.025$ &$0.043$ \\
NGC1404 &$-0.068$ &$-0.055$ &$0.86$ &$-0.18$ &$0.36$ &$2.0$ &$0.77$ &$1.3$ &$0.35$ &$-0.011$ &$0.021$\\
NGC1553 &$-0.18$ &$-0.18$ &$0.48$ &$0.14$ &$0.20$ &$0.14$ &$0.74$ &$0.91$ &$0.024$ &$5.0\times 10^{-3}$ &$7.4\times 10^{-3}$ \\
NGC3115 &$-0.022$ &$-0.039$ &$0.36$ &$-0.46$ &$-0.14$ &$0.23$ &$0.29$ &$0.32$ &$0.040$ &$-0.040$ &$-0.012$ \\
NGC3585 &$-0.095$ &$-0.11$ &$0.44$ &$-0.42$ &$-0.23$ &$0.15$ &$0.31$ &$0.47$ &$0.026$ &$-0.014$ &$-7.7\times 10^{-3}$ \\
NGC3607 &$-0.016$ &$-7.6\times 10^{-4}$ &$0.93$ &$0.29$ &$0.61$ &$0.43$ &$0.21$ &$0.18$ &$0.074$ &$0.012$ &$0.025$ \\
NGC4125 &\ldots &$-0.089$ &$0.37$ &\ldots &$0.27$ &$0.53$ &\ldots &$0.61$ &$0.023$ &\ldots &$8.5\times 10^{-3}$ \\
NGC4261 &$-0.11$ &$-0.032$ &$1.5$ &$3.7$ &$1.7$ &$2.0$ &$2.6$ &$-0.023$ &$0.34$ &$0.11$ &$0.050$ \\
NGC4365 &$-0.043$ &$-0.016$ &$0.86$ &$0.044$ &$0.80$ &$0.61$ &$0.59$ &$0.59$ &$0.11$ &$1.9\times 10^{-3}$ &$0.035$ \\
NGC4472 &$-0.035$ &$-0.042$ &$0.69$ &$-1.3$ &$0.051$ &$0.16$ &$0.55$ &$0.30$ &$0.028$ &$-0.034$ &$1.3\times 10^{-3}$ \\
NGC4494 &$-3.5\times 10^{-3}$ &$-0.085$ &$0.42$ &$-0.32$ &$0.95$ &$0.42$ &$0.41$ &$0.053$ &$0.071$ &$-0.022$ &$0.065$ \\
NGC4552 &$-0.032$ &$-0.060$ &$0.51$ &$-0.61$ &$0.66$ &$0.57$ &$0.95$ &$0.20$ &$0.098$ &$-0.046$ &$0.051$ \\
NGC4621 &$0.011$ &$-0.014$ &$0.51$ &$0.87$ &$1.8$ &$0.40$ &$-0.039$ &$-0.066$ &$0.069$ &$0.051$ &$0.11$ \\
NGC4649 &$-0.017$ &$-0.014$ &$0.68$ &$0.22$ &$1.5$ &$0.25$ &$0.19$ &$0.23$ &$0.041$ &$7.6\times 10^{-3}$ &$0.050$ \\
NGC5018 &$-0.011$ &$0.097$ &$1.8$ &$1.2$ &$0.043$ &$0.22$ &$0.10$ &$-0.13$ &$0.038$ &$0.013$ &$4.4\times 10^{-4}$ \\
NGC5845 &\ldots &$-0.019$ &$0.59$ &\ldots &$0.59$ &$0.43$ &\ldots &$7.4\times 10^{-3}$ &$0.073$ &\ldots &$0.045$ \\
NGC5846 &$-0.014$ &$-0.048$ &$0.57$ &$-0.74$ &$-0.090$ &$0.62$ &$0.64$ &$0.61$ &$0.088$ &$-0.029$ &$-3.5\times 10^{-3}$
\enddata
\tablecomments{Error-budget for \gcinc, \lgc, \Sn\ and \Sl, showing the statistical errors 
($\Delta_{stat}$) and an estimate of the impact on the parameters from allowing the GCLF turnover
to fit freely while fitting all of the available data ($\Delta_{free}$)  
or from adopting a GCLF width of 1.5~mag ($\Delta_{wid}$); see text.
We indicate by ellipses cases where the data could not be constrained (NGC\thin 1387 and NGC\thin 5845) or where the parameters were already derived allowing the GCLF turnover to be 
free (NGC\thin 4125).}
\end{deluxetable*}

\subsection{Source Detection Incompleteness} \label{sect_incompleteness}
In order to measure accurately the GCLF in each galaxy it is essential to take into
account the source detection incompleteness.
The ability of the algorithm outlined in \S~\ref{sect_reduction} to detect point-sources
is a strong function both of the source flux and the ``background'' level, which varies
dramatically over the WFPC2 field of view. It is therefore certain that some 
fraction of the GCs in any given galaxy will not be detected. To estimate the completeness
of our source lists, we therefore carried out extensive Monte-Carlo
simulations. Starting with the V-magnitude (B-band for NGC\thin 1404 and NGC\thin 1399)
of the faintest sources detected in 
each galaxy, we stepped through a series of magnitude bins, each separated by 0.2~mag, 
in each of which we performed 10 Monte-Carlo simulations to estimate the 
completeness level. 
We have explicitly confirmed that starting from one magnitude fainter than this limit
makes no measurable difference to our key results for one galaxy, NGC\thin 4649.
Each simulation entailed the addition of ten simulated point 
sources to the actual WFPC2 images of the galaxy in each filter. The V-I
colours of the sources were chosen randomly by drawing, with replacement, from the 
V-I distribution of the detected sources. The source detection
algorithm was run on each image and the number of simulated sources which were detected,
as well as their measured magnitudes, was determined. Once 98 percent of the simulated
sources were detected in any bin, the simulations were stopped and all sources
brighter than the corresponding magnitude were assumed detected at their true
magnitude (this is, in general, a very good approximation since the measurement errors
on the brightest GCs are typically much less than the 0.2~mag binsize we adopt).
The sources were assumed to be distributed as the optical light, which,
for these purposes was modelled as an elliptical de Vaucouleurs profile centred at the 
peak of the galaxy emission, with the size and orientation given in 
\citet[][except for NGC\thin 1332, for which the position angle was modified 
to match better the observed image of the galaxy]{devaucouleurs91}.
Point-source images were generated with the \tinytim\ tool for the appropriate filter
and focal plane position of the source, assuming the spectrum of a K4V star.

Any given parameterization of the GCLF which we fit
predicts the number of sources which should, in the idealized case, 
be observed in any given magnitude bin. To account for incompleteness
and statistical errors, this model was convolved, in each bin, with a function,
determined from our simulations, which redistributes the sources into adjacent bins, 
based on the likelihood it is detected in each magnitude bin.
The corrected model can then be fitted directly
to the data. This method is exactly analogous to computing
a ``response matrix'', through which any parameterized model 
can be folded in X-ray spectral-fitting \citep[\eg][]{davis01a}. We discuss
this method, and the intrinsic uncertainty associated with our method of 
computing the matrix  in Appendix~\ref{sect_respmatrix}.
This procedure not only accounts for source detection incompleteness, but also
the Eddington bias, while preserving the statistical integrity of the data.

\subsection{Fitting the GCLFs} \label{sect_gclf_fits}
\begin{figure}
\centering
\includegraphics[scale=0.35]{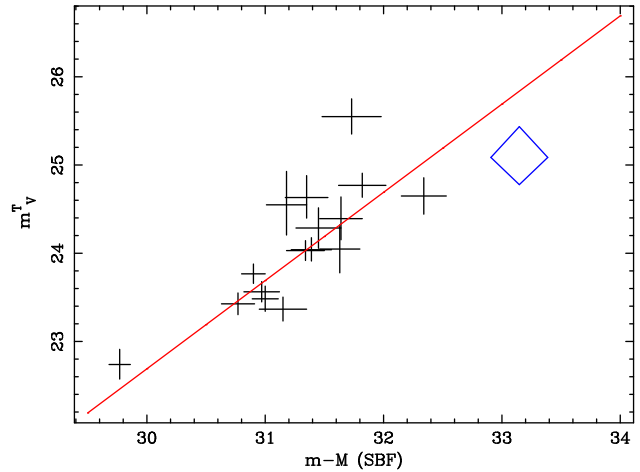}
\caption{Comparison of the measured GCLF turnover and the corrected I-band
SBF distance moduli (Table~\ref{table_obs}).
NGC\thin 5018, for which the distance was derived from the 
${\rm D_n}$-$\sigma$ relation \citep{faber89}, is shown as a diamond. 
The straight line denotes the best-fitting straight line
relation (see text), which is consistent with previously-reported correlations.
This demonstrates the utility of the GCLF turnover magnitude as an approximate
``standard candle''.} \label{fig_compare_distmod}
\end{figure}
For each galaxy, we accumulated a V-band (B-band for NGC\thin 1399 and 
NGC\thin 1404) GCLF by constructing a histogram of the numbers of sources observed
in a series of magnitude bins of width 0.2~mag. We created an associated response
matrix, as described in Appendix~\ref{sect_respmatrix} and fitted the data with
dedicated software through a Cash-C statistic function minimization procedure. 
We modelled the GC magnitude distribution as a simple Gaussian, \ie
\begin{equation}
\frac{d n_{GC}}{d V} = \frac{N_{TOT}}{\sqrt{2 \pi} \sigma_V} \exp
\left( -\frac{(V-V_T)^2}{2 \sigma_V^2}\right)
\end{equation}
where $n_{GC}$ is the number density of GCs, as a function of V-band
luminosity (V), $N_{TOT}$ is the total number of GCs and $V_T$ is the
apparent V-band peak luminosity (``GCLF turnover'').
This has been previously shown to be a reasonable
description of the GCLF shape \citep[\eg][]{harris91a,kundu01b}. For consistency
with past analysis, and to minimize the sensitivity of our results to
the our source detection incompleteness algorithm, we initially fitted only those 
data-bins for which the source detection completeness was estimated to be 50\% 
or greater. In general, our incompleteness cut allowed insufficient data below the 
GCLF turnover to constrain all three parameters of the Gaussian fit and so, 
based on past work \citep{kundu01b,jordan06a}, we fixed the GCLF half-width
($\sigma_V$) to 1.3~mag. This should be a reasonable average value for giant
elliptical galaxies, although in individual objects significantly broader or 
narrower GCLFs may exist.

In Table~\ref{table_derived} we list the V-band turnover magnitude, \mturn\
for each galaxy. To enable a consistent comparison, we convert B-band
turnover magnitudes for NGC\thin 1399 and NGC\thin 1404 to the V-band by assuming 
a constant B-V of 0.97, consistent
with the colours of these galaxies, and also consistent with a $\sim$solar
metallicity 10~Gyr stellar population \citep{maraston05a}. We ignored
NGC\thin 1387 and NGC\thin 5845, for which there were too few 
measured GCs to make the GCLF turnover measured in this way reliable.
Where our samples overlap, our measured \mturn\ were in good agreement 
with those found by \citet{kundu01b,kundu01} and \citet[][who fitted a slightly
different functional form to the GCLF]{larsen01a}, giving us confidence in
our fitting procedure.

In Fig~\ref{fig_compare_distmod} we show \mturn\ as a function of 
the distance modulus. Taking into account 
errors on both the x and y data values 
we fitted the data with a function of the form
${\rm m^T_V = (m-M)_{SBF} + M^T_V}$, where \Mturn, the absolute turnover
magnitude, was fitted freely. The fit was formally unacceptable 
($\chi^2$/dof=35/16), indicating significantly more scatter about this relation
than the statistical errors. Omitting the most discrepant data-point,
that of NGC\thin 4125, significantly improved the fit
($\chi^2$/dof=23/15), although it remained only marginally acceptable. For 
this fit we constrained \Mturn=$-7.30\pm0.05$, in good agreement with 
previous estimates
\citep[\eg][]{kundu01b}, especially when we account for recent revisions
to the Cepheid zero-point \citep[\eg][]{jensen03}.
In general, however, the {\em ad hoc} removal of data-points should
be avoided. Taking our fit to NGC\thin 4125 at face value implies 
intrinsic scatter (estimated as described in \S~\ref{sect_mass_metallicity}) 
of 0.4~mag, with a 
90\% lower limit of 0.26~mag, and a best-fitting \Mturn\ of $-7.26\pm 0.15$.

In order to compare the LMXB and GC populations in \lmxbpaper, we require an
estimate of the fraction of the GC luminosity which is detected in the WFPC2
field of view, \figc. To estimate this for each galaxy, we adopted the 
canonical GCLF (\Mturn=-7.3, $\sigma_V$=1.3) to each dataset (except for
NGC\thin 4125, for which we adopted \Mturn\ derived from fitting) and
computed the observed and total luminosity of GCs by 
appropriately integratingthe incompleteness-corrected and the uncorrected models.
We tabulate \figc\ in Table~\ref{table_derived}.

\subsection{Specific frequency and luminosity} \label{sect_gclf_individual}
\begin{figure*}
\centering
\includegraphics[scale=0.7]{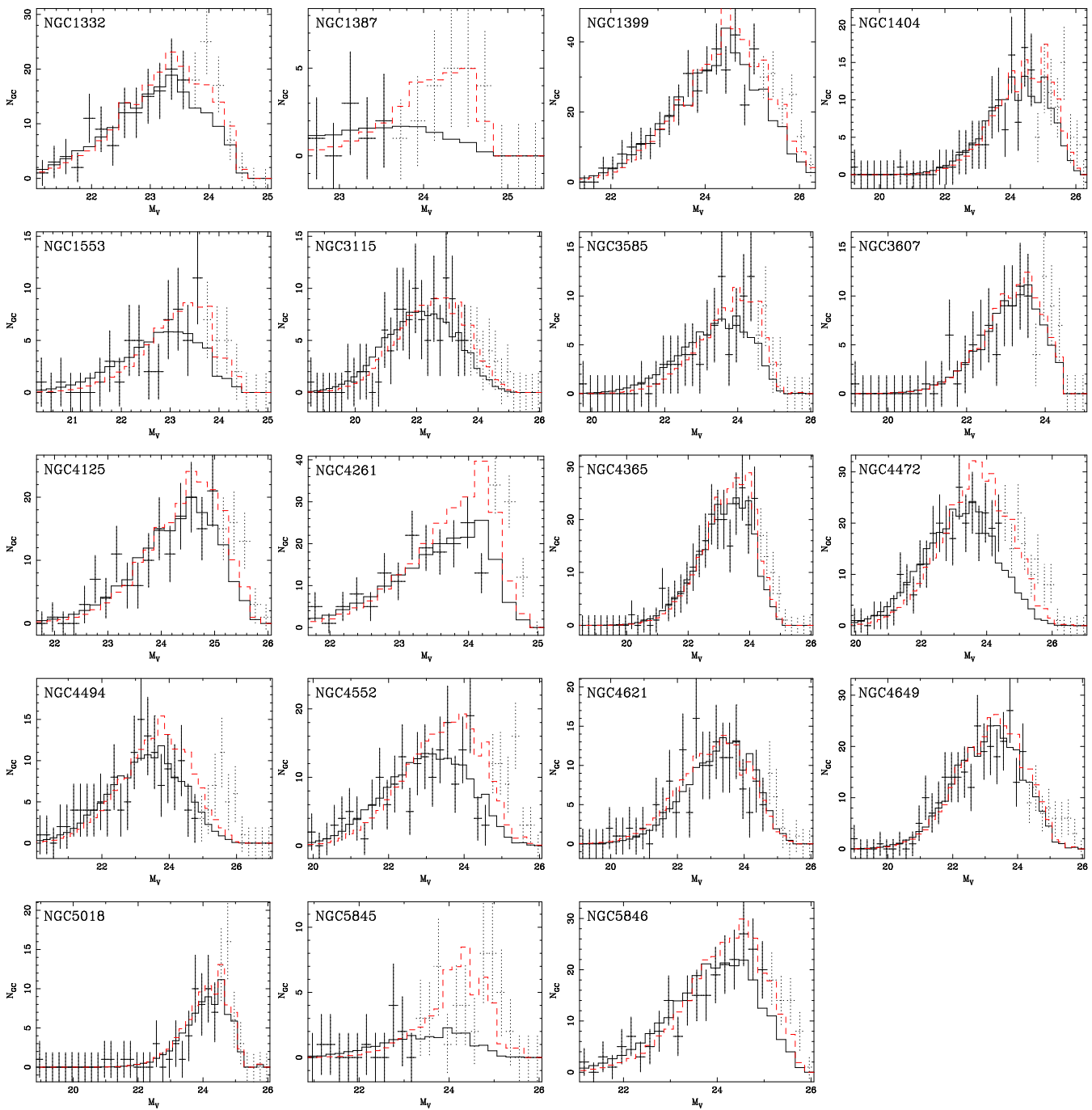}
\caption{GCLFs of each of the galaxies in the sample, shown with the 
best-fit Gaussian model (solid line), 
corrected for source detection incompleteness. Data-bins for which source
detection completeness was below 50\% (not included in the fit) are drawn with 
dotted lines. The dashed line shows the best-fitting model if all of the data
are fitted and the turnover magnitude is allowed to fit freely. Although the 
data were fitted by minimizing the Cash-C statistic, we show error-bars 
estimated from Gehrels' approximation (for the upper limit) to guide the eye. In most cases the 
data are well-represented by the simple Gaussian model, and the results are not
highly sensitive to the inclusion or exclusion of the more sources at the
uncertain faint end of the GCLF.} 
\label{fig_galaxy_gclfs}
\end{figure*}
To quantify the richness of the GC populations of each galaxy within the 
WFPC2 aperture, we computed two quantities, the
GC specific frequency, \Sn, and the GC specific luminosity, \Sl\
\citep[\eg][]{harris91a}, defined by the relations:
\begin{eqnarray}
S_N = N_{GC} 10^{0.4(M_V^{FOV}+15)}\\
S_L = 100 L_{GC}/L_V^{FOV}
\end{eqnarray}
where $N_{GC}$ is the total number of GCs inferred from fitting the GCLF, $L_{GC}$
is the total luminosity of the GCs inferred from the fit, $M_V^{FOV}$ is the 
V-band absolute magnitude of the galaxy in the field of view and $L_V^{FOV}$ is the 
corresponding luminosity.
Although we find significant intrinsic scatter in \Mturn, in order to produce
interesting constraints, we computed $N_{GC}$ and $L_{GC}$ having fixed \Mturn\
to -7.3 for each galaxy, except for NGC\thin 4125, for which we allowed the turnover
to fit freely since it is substantially fainter than \Mturn\
(Fig~\ref{fig_compare_distmod}). We estimated $M_V^{FOV}$, following
\citet{kundu01b}, by integrating the total counts in the field of view (having
excluded obvious bright interloper sources) and adopting a background level
of 0.01 and 0.052 $e^- s^{-1} pixel^{-1}$ for the PC and WF CCDs. We computed these
quantities in the V-band for all galaxies, excepting NGC\thin 1399\ and 
NGC\thin 1404, for which they were computed in the B-band. For these two systems,
we subsequently corrected \Sn\ to the V-band, assuming a galaxy B-V colour of 0.97.
The best-fitting results are shown in Table~\ref{table_derived}.
Where our data overlap, our computed \Sn\ (shown in Table~\ref{table_derived})
generally agree very well with those reported by \citet{kundu01b,kundu01}.
The GCLF data and the best-fitting models are shown in Fig~\ref{fig_galaxy_gclfs}.

This model fits the data brighter than the cut-off well. However,
extrapolation to fainter magnitudes generally reveals some discrepancies
(Fig~\ref{fig_galaxy_gclfs}). This may reflect actual differences between the GCLF
width and our canonical $\sigma_V$=1.3~mag, the increasing importance of 
(and difficulty
in eliminating) interlopers at faint magnitudes, and possible deficiencies in our
incompleteness-estimation algorithm. To investigate how important these effects
may be,  we experimented with fitting all of the 
data-points and freeing \mturn. 
The best-fitting models for such fits 
are shown (dashed lines) in Fig~\ref{fig_galaxy_gclfs}. 
In Table~\ref{table_syserr}, we tabulate the systematic uncertainties 
on \Sn\ and \Sl\ arising from this choice or, similarly,
fixing the GCLF width to $\sigma_V$=1.5 (see \S~\ref{sect_gclf_composite}).
Another potential source of systematic uncertainty is the assumed spatial 
distribution of the point sources in our simulations. In order to assess
how significantly this may affect our results, we recomputed the incompleteness
correction for one system, NGC\thin 4649, but distributing the simulated 
sources as a $\beta$ model, with $\beta=0.46$ and core radius 0.81\arcmin.
Such a model approximately fits the distribution of the LMXBs in this system.
We found \Sn\ changed by 0.20 and \Sl\ by 0.026, which are consistent with
the statistical errors, and other sources of systematic uncertainty for this
object.
In general, since \Sl\ is more sensitive to the 
bright end of the GCLF distribution, we found it was  more robust than \Sn\
to these choices; the systematic effect on \Sn\ of freeing \Mturn\
was generally larger than the statistical errors, whereas the 
impact on \Sl, while not negligible, was typically smaller than, or comparable
to, the statistical errors. \Sl\ also has the advantage that it is independent 
of the any error in the assumed distance to the galaxy.

\subsection{Composite GCLF} \label{sect_gclf_composite}
\begin{figure}
\centering
\includegraphics[scale=0.35]{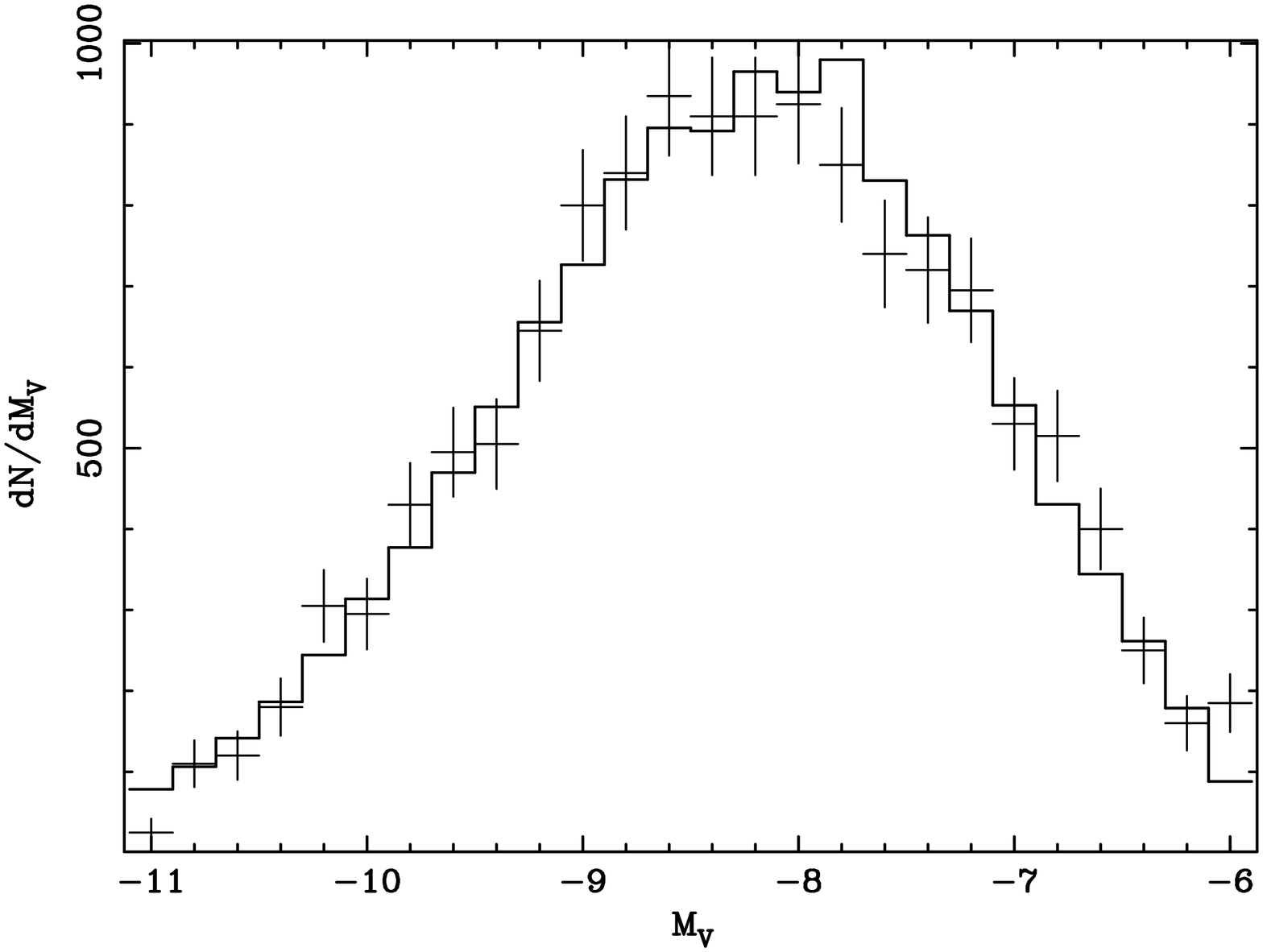}
\caption{Composite GCLF for all the galaxies in the sample. Also shown
is the best-fitting single Gaussian model, corrected for source detection
incompleteness. Although the data are fitted with the Cash-C statistic,
to guide the eye  we show error-bars estimated from the 
Gehrels' approximation. This demonstrates the good quality of a Gaussian GCLF
fit to the data.} \label{fig_composite_gclf}
\end{figure}
In \lmxbpaper\ we investigate the dependence of LMXBs in GCs by comparing the 
composite luminosity function of the GCs in all galaxies with that of the GCs containing
LMXBs.
In order to investigate the composite GCLF, we first computed V-band GCLFs for each 
galaxy in a set of standard absolute magnitude bins, and appropriate response 
files, as described in \S~\ref{sect_incompleteness}. To mitigate against the intrinsic
scatter in \Mturn, for each galaxy we adopted a distance modulus given by the
measured \mturn+7.3 to convert from apparent to absolute magnitude.
To take account of incompleteness effects the response matrices were averaged 
using  the \heasoft\ tool {\em addrmf}, weighting each matrix by the number of 
detected sources. To enable us to combine the GCLFs for NGC\thin 1399 and 
NGC\thin 1404 with the other data, we converted from the B to V band by assuming a 
constant B-V=0.97.

The combined GCLF was  well-fitted by a single Gaussian model, with 
$\sigma_V=1.51\pm0.04$. The data and best-fitting model are shown in 
Fig~\ref{fig_composite_gclf}. We note that the best-fitting $\sigma_V$ is slightly
larger than our adopted 1.3~mag used in computing \Sn\ and \Sl. This most likely
arises due to the GCLFs of some, although not necessarily all,
 of the objects being slightly broader than we
have assumed. The nature of a composite GCLF is that its shape is heavily biased towards
the most GC-rich galaxies, and so $\sigma_V=$1.5 need not be representative of the
majority of the galaxies in our sample. \citet{kundu01b} found that NGC\thin 4472
and NGC\thin 4649, which account for $\sim$20\% of the detected GCs, 
may have $\sigma_V$\gtsim1.5. 
The GCLF-determined distance moduli have 
statistical uncertainties typically of the order $\sim$0.1--0.2 (1-$\sigma$),
which should also cause the measured composite GCLF to be slightly broadened.
Nonetheless, we estimate in Table~\ref{table_syserr} the systematic effect on
the best-fitting \Sn\ and \Sl\ of fixing $\sigma_V$=1.5.

\section{Spatial distribution of Globular Clusters}
\begin{figure*}
\centering
\includegraphics[scale=0.7]{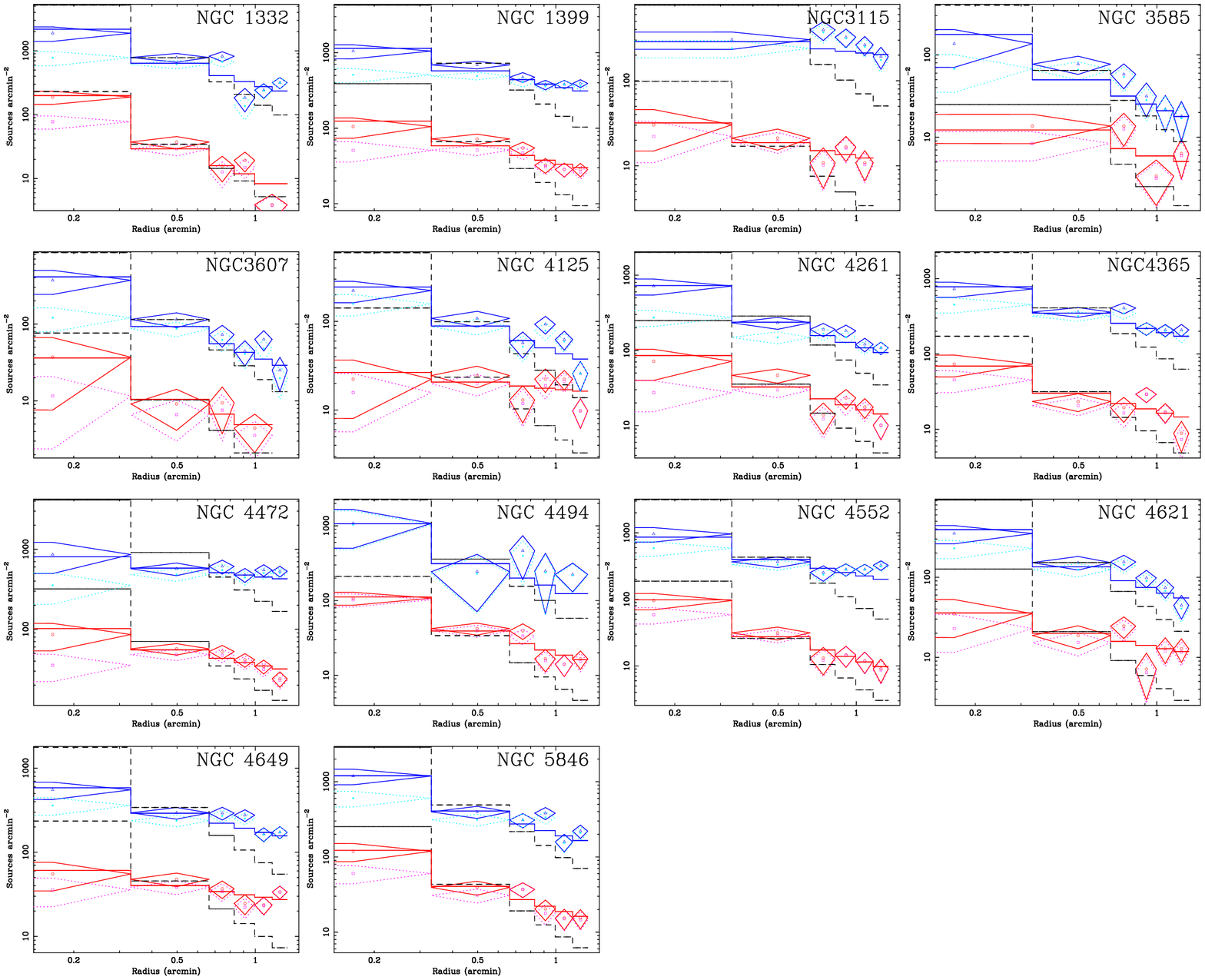}
\caption{Azimuthally-averaged radial  distribution of the GCs 
in a sub-sample of the galaxies. The radial distribution of the 
blue GCs are displaced upwards by an arbitrary amount for clarity.
The measured radial profiles data-points (solid lines) have been 
corrected for source detection incompleteness. We also show these
data without correction (dotted lines). In addition, we show 
(dashed line; black) the distribution of K-band optical light, 
and the best-fit simple powerlaw models to each dataset. In general,
the GC distribution is flatter than the optical light, and the blue GCs tend to
have a flatter distribution than the red GCs.} \label{fig_spatdist} 
\end{figure*}
\begin{deluxetable}{lll}
\tablecaption{Radial distribution fits \label{table_spatdist}}
\tablehead{
\colhead{Galaxy }&\colhead{$\alpha_{Red}$ }&\colhead{$\alpha_{Blue}$ }}
\startdata
NGC1332 &$-1.5^{+0.2}_{-0.1}$ &$-0.98^{+0.21}_{-0.18}$ \\
NGC1399 &$-0.73\pm 0.23$ &$-0.59\pm 0.17$ \\
NGC3115 &$-0.54^{+0.44}_{-0.36}$ &$-0.088^{+0.45}_{-0.38}$ \\
NGC3585 &$-0.71^{+0.73}_{-0.52}$ &$-0.94^{+0.34}_{-0.28}$ \\
NGC3607 &$-1.1^{+0.6}_{-0.4}$ &$-1.1\pm 0.2$ \\
NGC4125 &$-0.26^{+0.41}_{-0.35}$ &$-0.81\pm 0.23$ \\
NGC4261 &$-0.90^{+0.30}_{-0.25}$ &$-0.91^{+0.20}_{-0.18}$ \\
NGC4365 &$-0.79\pm 0.29$ &$-0.66^{+0.20}_{-0.18}$ \\
NGC4472 &$-0.61\pm 0.24$ &$-0.21\pm 0.27$ \\
NGC4494 &$-0.97\pm 0.18$ &$-0.82^{+0.47}_{-0.38}$ \\
NGC4552 &$-1.1\pm 0.2$ &$-0.62\pm 0.23$ \\
NGC4621 &$-0.58^{+0.40}_{-0.33}$ &$-0.85\pm 0.22$ \\
NGC4649 &$-0.42\pm 0.28$ &$-0.58\pm 0.19$ \\
NGC5846 &$-1.0\pm 0.2$ &$-0.87^{+0.21}_{-0.18}$ \\
\enddata
\tablecomments{Results from a power-law fit to the radial distribution
of the GCs within a radius of 1.5\arcmin. The power-law index, $\alpha$ is given 
separately for the red and blue GCs. The distribution of blue GCs tends 
to be flatter than for red GCs.}
\end{deluxetable}
In order to investigate the spatial distribution of the GCs, we
separately binned radial distributions for the red and blue GCs,
defining any cluster with V-I$>$1.1 as red, and the remaining clusters
as blue. We note that this does not cleanly separate the blue and red
sub-populations of clusters and so measured differences in the 
spatial distributions probably underestimate
intrinsic differences between the red and blue sub-populations. 
However, it does divide clusters on the 
basis of their {\em actual} metallicity (rather than asking to which
subpopulation they belong), thereby allowing us to assess whether,
for our galaxies, more metal-rich clusters are, on average, distributed
differently to more metal-poor ones. This distinction is extremely
important when comparing to the spatial distribution of LMXBs 
(\lmxbpaper).

To compute the radial profile, the radial distances of each source
from the optical centre of the galaxy, identified by eye, 
were binned into a set of 7 radial bins, with outer annuli
0.33, 0.66, 0.82, 1.0, 1.2, 1.3 and 1.5\arcmin, respectively. 
The data were rebinned to ensure at least 3 sources per bin
and the radial distribution was fitted with a variety of models
by a Cash-C statistic minimization algorithm.
To correct the spatial distributions of sources for detection incompleteness,
we computed a GCLF correction function, similar to that described in
\S~\ref{sect_incompleteness} for each bin.
Assuming the GCLF does not vary spatially, this correction was combined
with the measured GCLF for that galaxy to estimate the fraction of sources
which would be detected in each bin.

We show the spatial distributions of a sub-sample of the galaxies,
with sufficient measured GCs to yield interesting constraints,
in Fig~\ref{fig_spatdist}. We compare in this figure these distributions
to  the optical
light, as parameterized by a de Vaucouleurs model, with its effective
radius fixed to that measured in the K-band from the 2MASS Extended 
Source Catalog \citep{jarrett00a}.
We found that both the red and blue
GC distributions were significantly flatter than the optical light.
These data could be fitted adequately by a single powerlaw model,
\begin{equation}
\frac{d n_{GC}}{d r} = N\ r^{1+\alpha}
\end{equation}
where $dn_{GC}/dr$ is the number density of GCs as a function of radius, r,
and $\alpha$ is a parameter of the fit, as is the normalization $N$.
The results of the fits are tabulated in 
Table~\ref{table_spatdist} and shown in Fig~\ref{fig_spatdist}.
The GC distributions are considerably flatter 
than typically reported for studies with larger radial coverage
\cite[\eg][and references therein]{kisslerpatig97a}, but this simply 
reflects the flattening of the GC distributions in the galaxy 
core \citep[\eg][]{bassino06a}.
In general, the red GCs are significantly more 
centrally-peaked than the blue GCs, although there were a few exceptions to this
trend. This is probably a result of the size of the error-bars, 
the lack of full
azimuthal coverage,  the relatively small numbers of sources detected
and, critically, the small radial range. 
In any case, on average, the index for the blue GC
radial distribution is less negative by $\sim$0.15.

\section{Discussion}
In order to facilitate a direct comparison between LMXBs in early-type 
galaxies and possible GC candidates, we have presented a catalogue of GC
candidates and key derived properties based on the \hst\ WFPC2 centrally-pointed
observations of 19 nearby early-type galaxies. The complete source catalogue,
including coordinates and photometry, is given in Appendix~\ref{sect_sourcelist}. The total fraction
of the GC light detected and GC specific luminosities are given for each galaxy
in Table~\ref{table_derived}. In \lmxbpaper\ we combine these data with 
archival \chandra\ observations of early-type galaxies to show that the majority
of LMXBs must have formed in GCs.

We self-consistently derived various properties of the GC populations, including
fits to the GCLFs. We report specific frequency, \Sn, and specific luminosity,
\Sl, of the GC populations. Since the luminosity is dominated by clusters
brighter than the turnover, we found that  \Sl\ is much less sensitive than \Sn\
to uncertainties in the GCLF turnover or width, making it a more robust 
indicator of the richness of the GC population where 
a significant fraction of the GCs are undetected due to source
detection incompleteness. We find that the absolute magnitude of the 
GCLF turnover (\Mturn) exhibits intrinsic scatter from galaxy to galaxy
of $\sim$0.3--0.4~mag, when compared to SBF distance moduli, 
consistent with the estimate of \citet{ferrarese00a}.
\citet{kundu01b} argued the intrinsic uncertainty in the GCLF distance measure is 
$\sim0.14$~mag, but only when using a small subset of the galaxies in their sample
with the smallest error-bars on \mturn. However, there is a significant correlation
between (\mturn-\mturn$_{\rm model}$) and the error on \mturn\
(as expected, since, the fainter the turnover, the less complete the data), 
and so this systematically selects against galaxies with faint \Mturn,
possibly reducing the scatter. Provided their error-estimates are reasonably
representative, one can adopt the procedure outlined in \S~\ref{sect_mass_metallicity} 
to estimate the intrinsic scatter for the  25 galaxies in their full sample which 
overlap the  SBF sample of \citet{tonry01}. Using this approach, we estimate 
an intrinsic scatter of $\sim$0.38~mag (with a 90\% 
lower limit of 0.23~mag), consistent with our results.
\citet{jordan06a} \citep[see also][]{jordan07a} fitted Gaussian models to the GCLFs of early-type galaxies
in Virgo, and found a trend of increasingly faint turnover magnitude with
fainter $M_B$. For their galaxies with $M_B$\ltsim -20, we estimate by eye an
intrinsic scatter of $\sim$0.3~mag, also consistent with the estimate from
our data.
To some extent, this scatter is driven by the presence of significant populations
of diffuse star clusters in two systems; it may be that the presence of 
similar populations in the galaxies we consider here have contributed to the 
significant scatter we observe.

In keeping with past studies of GCs using V and I-band photometry 
we did not find any evidence of a significant correlation between colour and
luminosity.  Converting the colour and magnitude of each GC to mass and metallicity,
however, we found evidence of a weak mass-metallicity correlation for the GC 
populations as a whole. This trend cannot simply be attributed to observational
biases, such as the systematic failure to detect faint (low-mass), red (metal-rich)
GCs, since that would actually result in a stronger correlation in the 
colour-magnitude diagram. Similarly, our assumptions of a constant age for all
GCs or a blue horizontal branch for all GCs do not appear to bias our results.

The apparent contradiction of no colour-magnitude correlation implying a 
mass-metallicity relation can be understood in terms of the weak 
dependence of the SSP mass-to-light ratio on the metallicity. 
To put this another
way, if there was no correlation between mass and metallicity, due to the dependence
of the mass-to-light ratio on metallicity, we would expect there to be a significant
colour-magnitude correlation. To investigate this further,
we  simulated an artificial dataset comprising a similar number of GCs to that
observed.
For each artificial GC, we randomly assigned a mass and metallicity value from,
respectively, uncorrelated log-normal and normal distributions 
which approximately matched the observed data. 
Using the appropriate V and I-band mass-to-light ratios from \citet{maraston05a}, 
these data were converted to V and I-band photometry,
and estimated experimental noise was added. 
There was strong evidence of a weak correlation (the probability of no correlation
was $<10^{-40}$) between colour and magnitude, with the fainter GCs being, 
on average redder. Conversely, we have also used similar simulations to generate
colour-magnitude diagrams from artificial GC datasets which obey the observed
mass-metallicity dependence. Exactly as expected, we find that there
is only weak evidence of a correlation in the resulting data (prob of 
no correlation $\sim$0.1\%, as compared to $\sim 10^{-37}$ in colour-metallicity
space), demonstrating that the metallicity-dependent mass-to-light ratio
can, indeed, destroy a colour-magnitude relation.

The galaxies in our sample span a range of magnitudes and so it is important
to consider whether the known correlation between the galaxy mass and the 
peak of the GC colour distribution may affect our results.
In part the relation reflects the increasing importance of red clusters in
higher-mass galaxies \citep[\eg][]{peng06a} and it is difficult to imagine how
that could introduce a spurious GC mass-metallicity correlation if one is not 
already present. More problematically though, \citeauthor{peng06a} report a correlation between 
the mean metallicity of the {\em blue} GCs and the galaxy mass and a similar
trend for the red clusters. However, lower-mass galaxies are observed to have systematically
narrower and fainter GCLFs \citep{jordan07a}, which would be consistent with
a trend for brighter GCs to be redder.
In any case, for the magnitude range of the galaxies we consider here, the scatter
about each correlation is at least as large as the trend.

After this work was submitted for publication, we became aware of a paper by
\citet{kundu08a} which suggests that an apparent mass-metallicity relation
may arise due to observational biases. He proposes that a weak mass-radius relationship,
coupled with differing errors in the photometric aperture correction (since 
larger clusters may be marginally resolved) in different wavebands may lead to
a spurious correlation. To investigate whether this could plausibly explain our
observations, we first estimated the error in
the aperture correction for both V and I band by convolving King models with varying
half light radii (\rh) with PSF images (as described in \S~\ref{sect_reduction}). Since
the vast majority of the GCs in our study were detected on the WF CCDs and in galaxies
more distant than 15~Mpc, we computed the correction for our WF photometry only and
adopted a 15~Mpc distance while allowing \rh\ to vary from 
1 to 8~pc \citep[consistent with observations, \eg][]{spitler06a}. Although the size of
the cluster can affect the photometry in either band by as much as 
$\sim$0.3~mag, the effect was 
very similar in each filter so that V-I only changed by $\ll 0.1$~mag. Such a small 
effect is not enough to wipe out the expected V-I {\em versus} V correlation in the case
that mass and metallicity are uncorrelated. We have verified this by adding a colour-size 
relation into our Monte-Carlo simulations detailed above. We adopted a pathological
monotonic linear relation between GC radius and V-band magnitude, which is much stronger
than the weak trend observed. For our simulated colour-magnitude diagram under the 
assumption of no mass-metallicity relation, we found that, although the mass-radius
effect slightly reduces the resulting colour-magnitude correlation, the probability
the data are consistent with no correlation is still $\sim 10^{-32}$.

While our results depend to some degree on the adopted SSP models, it is clear
that two independent sets of models--- those of \citet{maraston05a} and those
of \citet{fioc97a} both imply a significant correlation between the mass and 
metallicity in these objects. The different normalizations and slopes for
these models simply arise from differences in the predicted relations between
the V and I-band mass-to-light ratios and the metallicity between the sets
of models. However, {\em any} prediction of a metallicity-dependent 
mass-to-light ratio will clearly give rise to a mass-metallicity correlation
of this kind. 

It is also interesting to compare our best-fitting mass-metallicity relation with
that observed in early-type galaxies. We show in Fig~\ref{fig_composite_cmd}
an extrapolation of the best-fitting straight-line relation for early-type 
galaxies found by \citet{thomas05a}. Intriguingly, we find a reasonable
qualitative agreement, although the \citeauthor{thomas05a} relation is slightly
less steep and has a slightly higher normalization. 
The observed trend for GCs may arise from a number of effects, such as 
self-enrichment \citep[\eg][]{mieske06a}, in which the deeper potential wells of 
more massive GCs are more able to hold onto metals ejected from the very first
generations of stars within them than in less-massive systems. Alternatively,
the mass of a forming GC may be related to the mass (and metallicity) of its
parent gas cloud \citep[\eg][]{harris06a}.
Whatever processes actually underlie the mass-metallicity relation, the remarkable
amount of intrinsic scatter (0.6~dex in metallicity) implies that 
stochastic processes actually dominate in determining the metallicity 
of an individual GC. Alternatively if there are, in fact, metal-rich and 
metal-poor sub-populations that cannot be distinguished in our data due to the
statistical noise, and which exhibit different relations, the measured scatter
will actually substantially overestimate the magnitude of such effects.

Recent (B,I)-band photometry, which allows better separation of metal-rich and 
metal-poor clusters, has revealed 
colour-magnitude correlations for the
metal-poor (blue) sub-population of GCs in some early-type galaxies, but no
clear trend for the red population. 
Transforming a composite colour-magnitude
diagram of three giant Virgo ellipticals onto the mass-metallicity
plane, \citet{mieske06a} found little evidence of bimodality, but a 
trend of increasing average metallicity with mass, broadly consistent with our observed
mass-metallicity dependence. 
Although bimodal colour distributions need not necessarily imply bimodal metallicity
distributions \citep[\eg][]{richtler06a}, it is unclear whether the absence of
bimodality in the mass-metallicity plane is simply an artefact of noise
introduced during the transformation from colour-magnitude space. In particular,
for our data, the error-bars on [Z/H] were typically so large that two putative
distinct populations may have been blurred together.
This is particularly relevant when we consider whether the ``blue tilt'' and,
in particular, the lack of a colour-magnitude relation for the metal-rich clusters
found by previous authors, are
consistent with our data. To assess this, we simulated a set of artificial
GC photometry data-points in the (g,z) photometric bands, with random z-band
absolute magnitudes ranging from -7 to -11 and 
colours obeying the ``blue tilt'' relation found for blue GCs by \citet{strader06a}.
Transforming the data to the mass-metallicity plane using the models of \citet{maraston05a},
the data were approximately consistent with a simple linear relation of the form 
[Z/H]=-4.43+0.50$\log_{10}(M/M_\odot)$. Performing the same analysis for the 
red GCs found by \citeauthor{strader06a}, which we approximate as having g-z colours
of 1.4, we found [Z/H]$\simeq$-0.22. For a population of both blue and red GCs, we
would expect the mean mass-metallicity relation to be simply a weighted average of these two
relations. The existing best-fit slope and intercept are actually reproduced fairly 
well (0.25 and -2.3, respectively) for a (reasonable) blue GC fraction of $\sim$50\%.
This excellent agreement with our results indicates that we may be observing the 
same trend as was observed in the (g,z) band, albeit the different trends for the 
red and blue populations are averaged together.

\appendix
\section{Fitting the GCLF with a Response Matrix} \label{sect_respmatrix}
Our method to fit the GCLF took account of both the source detection incompleteness
and the effects of the Eddington bias by adapting the ``response matrix'' formalism
from X-ray astronomy \citep[\eg][]{davis01a}.
\citet{wang04a} used a similar, but not identical, method to correct for the Eddington
bias when fitting luminosity functions of X-ray binaries.
For each galaxy we have a 
set of  measured GC magnitudes (and intrinsic errors on the measured magnitudes),
which we binned into an arbitrary set of bins. The resulting histogram can be
written as $D_i$ ($i=1$\ldots n). Consider a GC with an intrinsic magnitude $m$,
which we {\em measure} and assign to one of these bins. Due to statistical
fluctuations there is a finite possibility that it will be detected in any one
of the bins 1\ldots n (or not at all). We can therefore define $p_{i}(m)$ as the probability
that this object is detected in bin i. If the differential GCLF
can be written $dn/dm$, it follows that:
\[
<D_i> = \int_{-\infty}^{+\infty} dm\  p_i(m) \frac{dn}{dm}(m) 
\]
where the angle brackets denote the expectation. We can approximate this integral
as a sum:
\[
<D_i> \simeq  \sum_{j=1}^{N} p_{ij} \int_{m_{j-1}}^{m_j} dm\  \frac{dn}{dm}(m) \equiv \sum_{j=1}^{N} p_{ij} f_j 
\]
where $p_{ij} = p_i(m_j)$, and the spacing between each $m_j$ ($j=1$\ldots N)
is small enough to ensure that $p_i(m)$ does not vary strongly between them.
For any given parameterized GCLF, we therefore have the expected number of GCs in
each bin i=1\ldots n, allowing us to construct a likelihood function  after
\citet{cash79a}. Maximizing the likelihood function enables the parameters which
define the GCLF shape to be constrained by the data.

The probability $p_{ij}$ is, in principle, dependent on the measurement errors,
source confusion and the ``background'' level (which varies strongly across
the image) so its computation from first principles is extremely demanding.
Instead, we choose to estimate it by adding objects of known intrinsic magnitude
$m_k$ to the image and measuring them from the data. Since this is equivalent to 
setting $f_j=\delta_{jk}$, where $\delta_{jk}$ is the Kronecker delta function,
it follows that $<D_i>=p_{ik}$. For each magnitude $m_k$, we added 100 stars in 
this manner to the image (as described in \S~\ref{sect_incompleteness}), 
and assumed that the arithmetic mean of our {\em measured} 
$D_i$ recovered for those stars was a good 
approximation for the expectation in order to estimate $p_{ik}$.
The accuracy of this assumption is dependent on the number of simulations 
performed and so we expect some intrinsic inaccuracy in our measured probabilities
which, in turn, may affect the likelihood function and derived parameters.
It is therefore important to assess whether 100 simulations are sufficient
for our purposes.

To assess the accuracy of the matrices $p_{ij}$, we performed a set of  Monte-Carlo
simulations to determined how accurately a known GCLF can be recovered. We adopted
an arbitrary analytical expression 
for the incompleteness as a function of measured magnitude,
which was chosen approximately to match the behaviour found for a representative
galaxy in our sample, NGC\thin 4365. We stress that we do not have to match exactly
the properties of the galaxy, only approximately for our present purposes. Adopting
measurement errors consistent with the observed data, we were able to simulate an
``observed'' dataset, given a known input luminosity function. In this way, we were
also able to generate corresponding response matrices, following the procedure
described in \S~\ref{sect_incompleteness}. We generated matrices using 100 simulated
sources per magnitude bin, as in  \S~\ref{sect_incompleteness} but also generated a 
separate set of matrices using 10000 sources per magnitude bin. We subsequently 
refer to these as the standard and large matrices, respectively.

We carried out 100 Monte Carlo simulations, on each of which we generated both a 
dataset and corresponding response matrices,
binned identically to the real data. For the input GCLF, we assumed
\mturn $=24.0$, $\sigma_V=1.3$, consistent with NGC\thin 4365, and a total number of 
500 GCs. 
For each dataset and response matrix pair, we fitted the simulated data, allowing both
\mturn\ and the normalization to fit freely. 
The mean and standard error of the best-fitting
parameters, averaged over the simulations, can be used to assess the bias and statistical error
(which will also include a contribution from the error on the response matrix due to the 
finite number of simulated sources used to create it)
implicit in our analysis. Applying the standard matrices,  we obtained
best-fitting \mturn = $24.00\pm 0.12$ (1-$\sigma$ error) and a normalization of 
$504\pm 41$ (1-$\sigma$) GCs. The best-fitting values are sufficiently close to the true
value to indicate that the results are not biased, and the recovered errors are broadly
consistent with those found for the real data. Adopting the large matrices, we recovered
best-fitting values of  \mturn = $24.02\pm 0.11$ (1-$\sigma$ errors) and a normalization of 
$506\pm 36$ (1-$\sigma$) GCs. Since increasing the number of simulations used to generate
the matrices by a factor 100 has virtually no impact on the recovered parameters or error
bars, it follows that the standard matrices are sufficient to fit the GCLF. 

As another way to assess the accuracy of the matrices, we also 
performed a set of simulations for each dataset, in which the response matrix was
replaced with an artificial matrix, $p^{\prime}_{ij}$,
 which was derived from the real matrix, with 
approximately the expected level of statistical noise added (assuming the ``observed''
$p_{ij}$ is exactly correct), and the data refitted. 
We performed 25 such simulations for each object and
adopted the 1-$\sigma$ scatter of the best-fitting parameters from our Monte-Carlo 
simulations as an estimate of the uncertainty arising from our matrix calculation.
We computed $p^{\prime}_{ij}$ in the following manner. For each magnitude $m_j$, 
we drew 100 sources which were distributed amongst the histogram data-bins i=1\ldots n
randomly but following the probability distribution $p_{ij}$. Denoting the resulting data
histogram as $(D^{\prime}_i)_j$, clearly  $<(D^\prime_i)_j> = 100 p_{ij}$ (since there are
100 sources). The actual values of $(D^{\prime}_i)_j$ will be distributed about this expectation
due to statistical scatter, allowing us to define $p^{\prime}_{ij}=(D^{\prime}_i)_j/100$.
In practice, we found the uncertainty arising from the matrix computation
for \Sn, \Sl\ and \mturn\ to be \ltsim 10\%, \ltsim 4\% and \ltsim 1\%, respectively,
which are far smaller than any other source of error in our calculations for these
quantities.

\section{Source Lists and Photometry} \label{sect_sourcelist}
We show in Table~\ref{table_srclist} a list of the GC candidates detected
in each galaxy. For each source, we provide a unique identification number 
within that galaxy, as well as its right ascension and declination. We provide
the V and I-band photometry for each candidate GC as well as the mass and
metallicity, having converted them using the SSP models of \citet{maraston05a},
assuming an age of 13~Gyr and a blue horizontal branch. All errors are quoted at 
the 1-$\sigma$ confidence level. Note that [Z/H] is constrained to lie in the range
covered by the models, \ie\ -2.25 to 0.67.

\begin{deluxetable}{lllllll}
\tablecaption{Catalogue of GC candidates \label{table_srclist}}
\tablehead{
\colhead{Source }&\colhead{RA }&\colhead{dec }&\colhead{V }&\colhead{I }&\colhead{Mass }&\colhead{[Z/H] } \\
\colhead{ }&\colhead{ }&\colhead{ }&\colhead{mag }&\colhead{mag }&\colhead{$10^6$\msun}&\colhead{ }}
\startdata
\multicolumn{7}{c}{\bf NGC 1332}\\
$1$ &03h26m17.34s &$-21^\circ20^\prime25.5^{\prime\prime}$ &$21.93\pm 0.04$ &$20.79\pm 0.03$ &$2.6^{+0.5}_{-0.3}$ &$-0.23^{+0.15}_{-0.28}$ \\
$2$ &03h26m16.94s &$-21^\circ20^\prime16.7^{\prime\prime}$ &$23.9\pm 0.2$ &$22.9\pm 0.2$ &$0.28^{+0.27}_{-0.05}$ &$-1.2^{+0.6}_{-1.0}$ \\
$3$ &03h26m17.38s &$-21^\circ20^\prime18.4^{\prime\prime}$ &$23.4\pm 0.1$ &$21.99\pm 0.09$ &$1.11\pm 0.07$ &$0.67^{+0.010}_{-0.29}$ \\
$4$ &03h26m17.58s &$-21^\circ20^\prime18.9^{\prime\prime}$ &$22.64\pm 0.08$ &$21.36\pm 0.06$ &$2.08^{+0.09}_{-0.2}$ &$0.44^{+0.23}_{-0.16}$ \\
$5$ &03h26m17.82s &$-21^\circ20^\prime19.1^{\prime\prime}$ &$23.4\pm 0.1$ &$22.22\pm 0.09$ &$0.69\pm 0.15$ &$-0.22^{+0.77}_{-0.30}$ 
\enddata
\tablecomments{The complete version of this table is in the electronic edition of the Journal. The printed edition contains only a sample.}
\end{deluxetable}


\acknowledgements
We would like to thank David Buote, Fabio Gastaldello and Luca Zappacosta
for stimulating discussions. 
We would like to thank Marc Seigar and Aaron Barth  for helpful discussions
regarding the analysis of the \hst\ data.
Some of the data presented in this paper were obtained from the Multimission Archive at 
the Space Telescope Science Institute ({MAST}). 
STScI is operated by the Association of Universities for Research in Astronomy, Inc., 
under NASA contract NAS5-26555. 
This research has also made use of the 
NASA/IPAC Extragalactic Database (\ned)
which is operated by the Jet Propulsion Laboratory, California Institute of
Technology, under contract with NASA, and the HyperLEDA database
(http://leda.univ-lyon1.fr).
Support for this work was provided by NASA under grant 
NNG04GE76G issued through the Office of Space Sciences Long-Term
Space Astrophysics Program.

\bibliographystyle{apj_hyper}
\bibliography{paper_bibliography.bib}

\begin{thebibliography}{66}
\expandafter\ifx\csname natexlab\endcsname\relax\def\natexlab#1{#1}\fi

\bibitem[{{Bassino} {et~al.}(2006){Bassino}, {Richtler}, \&
  {Dirsch}}]{bassino06a}
\href{http://adsabs.harvard.edu/cgi-bin/nph-bib_query?bibcode=2006MNRAS.367..1%
56B&db_key=AST}{{Bassino}, L.~P., {Richtler}, T., \& {Dirsch}, B.} 2006,
  \mnras, 367, 156

\bibitem[{{Brodie} \& {Huchra}(1991)}]{brodie91a}
\href{http://adsabs.harvard.edu/cgi-bin/nph-bib_query?bibcode=1991ApJ...379..1%
57B&db_key=AST}{{Brodie}, J.~P. \& {Huchra}, J.~P.} 1991, \apj, 379, 157

\bibitem[{{Brodie} \& {Strader}(2006)}]{brodie06a}
\href{http://adsabs.harvard.edu/cgi-bin/nph-bib_query?bibcode=2006ARA\%26A..44%
..193B&db_key=AST}{{Brodie}, J.~P. \& {Strader}, J.} 2006, \araa, 44, 193

\bibitem[{{Cardelli} {et~al.}(1989){Cardelli}, {Clayton}, \&
  {Mathis}}]{cardelli89a}
\href{http://adsabs.harvard.edu/cgi-bin/nph-bib_query?bibcode=1989ApJ...345..2%
45C&db_key=AST}{{Cardelli}, J.~A., {Clayton}, G.~C., \& {Mathis}, J.~S.} 1989,
  \apj, 345, 245

\bibitem[{{Carpenter}(2001)}]{carpenter01a}
\href{http://adsabs.harvard.edu/cgi-bin/nph-bib_query?bibcode=2001AJ....121.28%
51C&db_key=AST}{{Carpenter}, J.~M.} 2001, \aj, 121, 2851

\bibitem[{{Cash}(1979)}]{cash79a}
\href{http://adsabs.harvard.edu/abs/1979ApJ...228..939C}{{Cash}, W.} 1979,
  \apj, 228, 939

\bibitem[{{Clark}(1975)}]{clark75a}
\href{http://adsabs.harvard.edu/cgi-bin/nph-bib_query?bibcode=1975ApJ...199L.1%
43C&db_key=AST}{{Clark}, G.~W.} 1975, \apjl, 199, L143

\bibitem[{{Davis}(2001)}]{davis01a}
\href{http://adsabs.harvard.edu/cgi-bin/nph-bib_query?bibcode=2001ApJ...548.10%
10D&db_key=AST}{{Davis}, J.~E.} 2001, \apj, 548, 1010

\bibitem[{{de Vaucouleurs} {et~al.}(1991){de Vaucouleurs}, {de Vaucouleurs},
  {Corwin}, {Buta}, {Paturel}, \& {Fouque}}]{devaucouleurs91}
\href{http://adsabs.harvard.edu/cgi-bin/nph-bib_query?bibcode=1991trcb.book...%
..D&amp;db_key=AST}{{de Vaucouleurs}, G., {de Vaucouleurs}, A., {Corwin},
  H.~G., {Buta}, R.~J., {Paturel}, G., \& {Fouque}, P.} 1991, {Third Reference
  Catalogue of Bright Galaxies} (Volume 1-3, XII, 2069 pp.~7 figs..~
  Springer-Verlag Berlin Heidelberg New York)

\bibitem[{{Faber} {et~al.}(1989){Faber}, {Wegner}, {Burstein}, {Davies},
  {Dressler}, {Lynden-Bell}, \& {Terlevich}}]{faber89}
\href{http://adsabs.harvard.edu/cgi-bin/nph-bib_query?bibcode=1989ApJS...69..7%
63F&db_key=AST}{{Faber}, S.~M., {Wegner}, G., {Burstein}, D., {Davies}, R.~L.,
  {Dressler}, A., {Lynden-Bell}, D., \& {Terlevich}, R.~J.} 1989, \apjs, 69,
  763

\bibitem[{{Fabian} {et~al.}(1975){Fabian}, {Pringle}, \& {Rees}}]{fabian75a}
\href{http://adsabs.harvard.edu/cgi-bin/nph-bib_query?bibcode=1975MNRAS.172P..%
15F&db_key=AST}{{Fabian}, A.~C., {Pringle}, J.~E., \& {Rees}, M.~J.} 1975,
  \mnras, 172, 15P

\bibitem[{{Ferrarese} {et~al.}(2000){Ferrarese}, {Mould}, {Kennicutt},
  {Huchra}, {Ford}, {Freedman}, {Stetson}, {Madore}, {Sakai}, {Gibson},
  {Graham}, {Hughes}, {Illingworth}, {Kelson}, {Macri}, {Sebo}, \&
  {Silbermann}}]{ferrarese00a}
\href{http://adsabs.harvard.edu/cgi-bin/nph-bib_query?bibcode=2000ApJ...529..7%
45F&db_key=AST}{{Ferrarese}, L., {et~al.}} 2000, \apj, 529, 745

\bibitem[{{Fioc} \& {Rocca-Volmerange}(1997)}]{fioc97a}
\href{http://adsabs.harvard.edu/abs/1997A\%26A...326..950F}{{Fioc}, M. \&
  {Rocca-Volmerange}, B.} 1997, \aap, 326, 950

\bibitem[{{Fioc} \& {Rocca-Volmerange}(1999)}]{fioc99a}
\href{http://adsabs.harvard.edu/cgi-bin/nph-bib_query?bibcode=1999astro.ph.121%
79F&db_key=PRE}{{Fioc}, M. \& {Rocca-Volmerange}, B.} 1999, astro-ph/9912179

\bibitem[{{Gebhardt} \& {Kissler-Patig}(1999)}]{gebhardt99a}
\href{http://adsabs.harvard.edu/cgi-bin/nph-bib_query?bibcode=1999AJ....118.15%
26G&db_key=AST}{{Gebhardt}, K. \& {Kissler-Patig}, M.} 1999, \aj, 118, 1526

\bibitem[{{Geisler} {et~al.}(1996){Geisler}, {Lee}, \& {Kim}}]{geisler96a}
\href{http://adsabs.harvard.edu/cgi-bin/nph-bib_query?bibcode=1996AJ....111.15%
29G&db_key=AST}{{Geisler}, D., {Lee}, M.~G., \& {Kim}, E.} 1996, \aj, 111, 1529

\bibitem[{{Grindlay}(1987)}]{grindlay87a}
\href{http://adsabs.harvard.edu/abs/1987IAUS..125..173G}{{Grindlay}, J.~E.}
  1987, in IAU Symposium, Vol. 125, The Origin and Evolution of Neutron Stars,
  ed. D.~J. {Helfand} \& J.-H. {Huang}, 173--184

\bibitem[{{Harris}(1991)}]{harris91a}
\href{http://adsabs.harvard.edu/cgi-bin/nph-bib_query?bibcode=1991ARA\%26A..29%
..543H&db_key=AST}{{Harris}, W.~E.} 1991, \araa, 29, 543

\bibitem[{{Harris} {et~al.}(2006){Harris}, {Whitmore}, {Karakla}, {Oko{\'n}},
  {Baum}, {Hanes}, \& {Kavelaars}}]{harris06a}
\href{http://adsabs.harvard.edu/cgi-bin/nph-bib_query?bibcode=2006ApJ...636...%
90H&db_key=AST}{{Harris}, W.~E., {Whitmore}, B.~C., {Karakla}, D., {Oko{\'n}},
  W., {Baum}, W.~A., {Hanes}, D.~A., \& {Kavelaars}, J.~J.} 2006, \apj, 636, 90

\bibitem[{{Holtzman} {et~al.}(1995{\natexlab{a}}){Holtzman}, {Burrows},
  {Casertano}, {Hester}, {Trauger}, {Watson}, \& {Worthey}}]{holtzman95}
\href{http://adsabs.harvard.edu/cgi-bin/nph-bib_query?bibcode=1995PASP..107.10%
65H&db_key=AST}{{Holtzman}, J.~A., {Burrows}, C.~J., {Casertano}, S., {Hester},
  J.~J., {Trauger}, J.~T., {Watson}, A.~M., \& {Worthey}, G.}
  1995{\natexlab{a}}, \pasp, 107, 1065

\bibitem[{{Holtzman} {et~al.}(1995{\natexlab{b}}){Holtzman}, {Hester},
  {Casertano}, {Trauger}, {Watson}, {Ballester}, {Burrows}, {Clarke}, {Crisp},
  {Evans}, {Gallagher}, {Griffiths}, {Hoessel}, {Matthews}, {Mould}, {Scowen},
  {Stapelfeldt}, \& {Westphal}}]{holtzman95b}
\href{http://adsabs.harvard.edu/cgi-bin/nph-bib_query?bibcode=1995PASP..107..1%
56H&db_key=AST}{{Holtzman}, J.~A., {et~al.}} 1995{\natexlab{b}}, \pasp, 107,
  156

\bibitem[{{Humphrey} \& {Buote}(2004)}]{humphrey04a}
\href{http://adsabs.harvard.edu/cgi-bin/nph-bib_query?bibcode=2004ApJ...612..8%
48H&amp;db_key=AST}{{Humphrey}, P.~J. \& {Buote}, D.~A.} 2004, \apj, 612, 848

\bibitem[{{Humphrey} \& {Buote}(2006)}]{humphrey05a}
\href{http://adsabs.harvard.edu/cgi-bin/nph-bib_query?bibcode=2006ApJ...639..1%
36H&db_key=AST}{{Humphrey}, P.~J. \& {Buote}, D.~A.} 2006, \apj, 639, 136

\bibitem[{{Humphrey} \& {Buote}(2008)}]{humphrey06b}
\href{http://arxiv.org/abs/astro-ph/0612058}{{Humphrey}, P.~J. \& {Buote},
  D.~A.} 2008, \apj, in press, astro-ph/0612058

\bibitem[{{Humphrey} {et~al.}(2006){Humphrey}, {Buote}, {Gastaldello},
  {Zappacosta}, {Bullock}, {Brighenti}, \& {Mathews}}]{humphrey06a}
\href{http://adsabs.harvard.edu/cgi-bin/nph-bib_query?bibcode=2006ApJ...646..8%
99H&db_key=AST}{{Humphrey}, P.~J., {Buote}, D.~A., {Gastaldello}, F.,
  {Zappacosta}, L., {Bullock}, J.~S., {Brighenti}, F., \& {Mathews}, W.~G.}
  2006, \apj, 646, 899

\bibitem[{{Irwin}(2005)}]{irwin05a}
\href{http://adsabs.harvard.edu/cgi-bin/nph-bib_query?bibcode=2005ApJ...631..5%
11I&db_key=AST}{{Irwin}, J.~A.} 2005, \apj, 631, 511

\bibitem[{{Ivanova}(2006)}]{ivanova06c}
\href{http://adsabs.harvard.edu/cgi-bin/nph-bib_query?bibcode=2006ApJ...636..9%
79I&db_key=AST}{{Ivanova}, N.} 2006, \apj, 636, 979

\bibitem[{{Jacoby} {et~al.}(1992){Jacoby}, {Branch}, {Clardullo}, {Davies},
  {Harris}, {Pierce}, {Pritchet}, {Tonry}, \& {Welch}}]{jacoby92}
\href{http://cdsads.u-strasbg.fr/cgi-bin/nph-bib_query?bibcode=1992PASP..104..%
599J&amp;db_key=AST}{{Jacoby}, G.~H., {Branch}, D., {Clardullo}, R., {Davies},
  R.~L., {Harris}, W.~E., {Pierce}, M.~J., {Pritchet}, C.~J., {Tonry}, J.~L.,
  \& {Welch}, D.~L.} 1992, \pasp, 104, 599

\bibitem[{{Jarrett}(2000)}]{jarrett00a}
\href{http://adsabs.harvard.edu/cgi-bin/nph-bib_query?bibcode=2000PASP..112.10%
08J&db_key=AST}{{Jarrett}, T.~H.} 2000, \pasp, 112, 1008

\bibitem[{{Jensen} {et~al.}(2003){Jensen}, {Tonry}, {Barris}, {Thompson},
  {Liu}, {Rieke}, {Ajhar}, \& {Blakeslee}}]{jensen03}
\href{http://adsabs.harvard.edu/cgi-bin/nph-bib_query?bibcode=2003ApJ...583..7%
12J&amp;db_key=AST}{{Jensen}, J.~B., {Tonry}, J.~L., {Barris}, B.~J.,
  {Thompson}, R.~I., {Liu}, M.~C., {Rieke}, M.~J., {Ajhar}, E.~A., \&
  {Blakeslee}, J.~P.} 2003, \apj, 583, 712

\bibitem[{{Jord{\'a}n} {et~al.}(2004){Jord{\'a}n}, {C{\^o}t{\'e}}, {Ferrarese},
  {Blakeslee}, {Mei}, {Merritt}, {Milosavljevi{\'c}}, {Peng}, {Tonry}, \&
  {West}}]{jordan04a}
\href{http://adsabs.harvard.edu/cgi-bin/nph-bib_query?bibcode=2004ApJ...613..2%
79J&db_key=AST}{{Jord{\'a}n}, A., {C{\^o}t{\'e}}, P., {Ferrarese}, L.,
  {Blakeslee}, J.~P., {Mei}, S., {Merritt}, D., {Milosavljevi{\'c}}, M.,
  {Peng}, E.~W., {Tonry}, J.~L., \& {West}, M.~J.} 2004, \apj, 613, 279

\bibitem[{{Jord{\'a}n} {et~al.}(2007){Jord{\'a}n}, {McLaughlin},
  {C{\^o}t{\'e}}, {Ferrarese}, {Peng}, {Mei}, {Villegas}, {Merritt}, {Tonry},
  \& {West}}]{jordan07a}
\href{http://adsabs.harvard.edu/abs/2007ApJS..171..101J}{{Jord{\'a}n}, A.,
  {McLaughlin}, D.~E., {C{\^o}t{\'e}}, P., {Ferrarese}, L., {Peng}, E.~W.,
  {Mei}, S., {Villegas}, D., {Merritt}, D., {Tonry}, J.~L., \& {West}, M.~J.}
  2007, \apjs, 171, 101

\bibitem[{{Jord{\'a}n} {et~al.}(2006){Jord{\'a}n}, {McLaughlin},
  {C{\^o}t{\'e}}, {Ferrarese}, {Peng}, {Blakeslee}, {Mei}, {Villegas},
  {Merritt}, {Tonry}, \& {West}}]{jordan06a}
\href{http://adsabs.harvard.edu/cgi-bin/nph-bib_query?bibcode=2006ApJ...651L..%
25J&db_key=AST}{{Jord{\'a}n}, A., {et~al.}} 2006, \apjl, 651, L25

\bibitem[{{Juett}(2005)}]{juett05a}
\href{http://adsabs.harvard.edu/cgi-bin/nph-bib_query?bibcode=2005ApJ...621L..%
25J&db_key=AST}{{Juett}, A.~M.} 2005, \apjl, 621, L25

\bibitem[{{Kim} {et~al.}(2006){Kim}, {Kim}, {Fabbiano}, {Lee}, {Park},
  {Geisler}, \& {Dirsch}}]{kim05a}
\href{http://adsabs.harvard.edu/cgi-bin/nph-bib_query?bibcode=2006ApJ...647..2%
76K&db_key=AST}{{Kim}, E., {Kim}, D.-W., {Fabbiano}, G., {Lee}, M.~G., {Park},
  H.~S., {Geisler}, D., \& {Dirsch}, B.} 2006, \apj, 647, 276

\bibitem[{{King}(1962)}]{king62a}
\href{http://adsabs.harvard.edu/abs/1962AJ.....67..471K}{{King}, I.} 1962, \aj,
  67, 471

\bibitem[{{Kissler-Patig}(1997)}]{kisslerpatig97a}
\href{http://adsabs.harvard.edu/cgi-bin/nph-bib_query?bibcode=1997A\%26A...319%
...83K&db_key=AST}{{Kissler-Patig}, M.} 1997, \aap, 319, 83

\bibitem[{{Kroupa}(2001)}]{kroupa01a}
\href{http://adsabs.harvard.edu/cgi-bin/nph-bib_query?bibcode=2001MNRAS.322..2%
31K&db_key=AST}{{Kroupa}, P.} 2001, \mnras, 322, 231

\bibitem[{{Kundu}(2008)}]{kundu08a}
\href{http://adsabs.harvard.edu/abs/2008AJ....136.1013K}{{Kundu}, A.} 2008,
  \aj, 136, 1013

\bibitem[{{Kundu} {et~al.}(2002){Kundu}, {Maccarone}, \& {Zepf}}]{kundu02a}
\href{http://adsabs.harvard.edu/cgi-bin/nph-bib_query?bibcode=2002ApJ...574L..%
.5K&db_key=AST}{{Kundu}, A., {Maccarone}, T.~J., \& {Zepf}, S.~E.} 2002, \apjl,
  574, L5

\bibitem[{{Kundu} \& {Whitmore}(2001{\natexlab{a}})}]{kundu01b}
\href{http://adsabs.harvard.edu/cgi-bin/nph-bib_query?bibcode=2001AJ....121.29%
50K&amp;db_key=AST}{{Kundu}, A. \& {Whitmore}, B.~C.} 2001{\natexlab{a}}, \aj,
  121, 2950

\bibitem[{{Kundu} \& {Whitmore}(2001{\natexlab{b}})}]{kundu01}
\href{http://adsabs.harvard.edu/cgi-bin/nph-bib_query?bibcode=2001AJ....122.12%
51K&amp;db_key=AST}{{Kundu}, A. \& {Whitmore}, B.~C.} 2001{\natexlab{b}}, \aj,
  122, 1251

\bibitem[{{Kundu} {et~al.}(1999){Kundu}, {Whitmore}, {Sparks}, {Macchetto},
  {Zepf}, \& {Ashman}}]{kundu99a}
\href{http://adsabs.harvard.edu/cgi-bin/nph-bib_query?bibcode=1999ApJ...513..7%
33K&db_key=AST}{{Kundu}, A., {Whitmore}, B.~C., {Sparks}, W.~B., {Macchetto},
  F.~D., {Zepf}, S.~E., \& {Ashman}, K.~M.} 1999, \apj, 513, 733

\bibitem[{{Larsen} {et~al.}(2001){Larsen}, {Brodie}, {Huchra}, {Forbes}, \&
  {Grillmair}}]{larsen01a}
\href{http://adsabs.harvard.edu/cgi-bin/nph-bib_query?bibcode=2001AJ....121.29%
74L&db_key=AST}{{Larsen}, S.~S., {Brodie}, J.~P., {Huchra}, J.~P., {Forbes},
  D.~A., \& {Grillmair}, C.~J.} 2001, \aj, 121, 2974

\bibitem[{{Maccarone} {et~al.}(2004){Maccarone}, {Kundu}, \&
  {Zepf}}]{maccarone04a}
\href{http://adsabs.harvard.edu/cgi-bin/nph-bib_query?bibcode=2004ApJ...606..4%
30M&db_key=AST}{{Maccarone}, T.~J., {Kundu}, A., \& {Zepf}, S.~E.} 2004, \apj,
  606, 430

\bibitem[{{Maraston}(1998)}]{maraston98a}
\href{http://adsabs.harvard.edu/cgi-bin/nph-bib_query?bibcode=1998MNRAS.300..8%
72M&db_key=AST}{{Maraston}, C.} 1998, \mnras, 300, 872

\bibitem[{{Maraston}(2005)}]{maraston05a}
\href{http://adsabs.harvard.edu/abs/2005MNRAS.362..799M}{{Maraston}, C.} 2005,
  \mnras, 362, 799

\bibitem[{{Mieske} {et~al.}(2006){Mieske}, {Jord{\'a}n}, {C{\^o}t{\'e}},
  {Kissler-Patig}, {Peng}, {Ferrarese}, {Blakeslee}, {Mei}, {Merritt}, {Tonry},
  \& {West}}]{mieske06a}
\href{http://adsabs.harvard.edu/abs/2006ApJ...653..193M}{{Mieske}, S.,
  {et~al.}} 2006, \apj, 653, 193

\bibitem[{{Peng} {et~al.}(2006){Peng}, {Jord{\'a}n}, {C{\^o}t{\'e}},
  {Blakeslee}, {Ferrarese}, {Mei}, {West}, {Merritt}, {Milosavljevi{\'c}}, \&
  {Tonry}}]{peng06a}
\href{http://adsabs.harvard.edu/cgi-bin/nph-bib_query?bibcode=2006ApJ...639...%
95P&db_key=AST}{{Peng}, E.~W., {Jord{\'a}n}, A., {C{\^o}t{\'e}}, P.,
  {Blakeslee}, J.~P., {Ferrarese}, L., {Mei}, S., {West}, M.~J., {Merritt}, D.,
  {Milosavljevi{\'c}}, M., \& {Tonry}, J.~L.} 2006, \apj, 639, 95

\bibitem[{{Press} {et~al.}(1992){Press}, {Teukolsky}, {Vetterling}, \&
  {Flannery}}]{nr}
\href{http://www.nr.com/}{{Press}, W.~H., {Teukolsky}, S.~A., {Vetterling},
  W.~T., \& {Flannery}, B.~P.} 1992, {Numerical Recipes in C: The Art of
  Scientific Computing. Second Ed.} (C.U.P.)

\bibitem[{{Richtler}(2006)}]{richtler06a}
\href{http://adsabs.harvard.edu/cgi-bin/nph-bib_query?bibcode=2006BASI...34...%
83R&db_key=AST}{{Richtler}, T.} 2006, Bulletin of the Astronomical Society of
  India, 34, 83

\bibitem[{{Sarazin} {et~al.}(2003){Sarazin}, {Kundu}, {Irwin}, {Sivakoff},
  {Blanton}, \& {Randall}}]{sarazin03}
\href{http://cdsads.u-strasbg.fr/cgi-bin/nph-bib_query?bibcode=2003ApJ...595..%
743S&amp;db_key=AST}{{Sarazin}, C.~L., {Kundu}, A., {Irwin}, J.~A., {Sivakoff},
  G.~R., {Blanton}, E.~L., \& {Randall}, S.~W.} 2003, \apj, 595, 743

\bibitem[{{Schlegel} {et~al.}(1998){Schlegel}, {Finkbeiner}, \&
  {Davis}}]{schlegel98a}
\href{http://adsabs.harvard.edu/cgi-bin/nph-bib_query?bibcode=1998ApJ...500..5%
25S&db_key=AST}{{Schlegel}, D.~J., {Finkbeiner}, D.~P., \& {Davis}, M.} 1998,
  \apj, 500, 525

\bibitem[{{Sivakoff} {et~al.}(2007){Sivakoff}, {Jord{\'a}n}, {Sarazin},
  {Blakeslee}, {C{\^o}t{\'e}}, {Ferrarese}, {Juett}, {Mei}, \&
  {Peng}}]{sivakoff06a}
\href{http://adsabs.harvard.edu/abs/2007ApJ...660.1246S}{{Sivakoff}, G.~R.,
  {Jord{\'a}n}, A., {Sarazin}, C.~L., {Blakeslee}, J.~P., {C{\^o}t{\'e}}, P.,
  {Ferrarese}, L., {Juett}, A.~M., {Mei}, S., \& {Peng}, E.~W.} 2007, \apj,
  660, 1246

\bibitem[{{Smits} {et~al.}(2006){Smits}, {Maccarone}, {Kundu}, \&
  {Zepf}}]{smits06a}
\href{http://adsabs.harvard.edu/cgi-bin/nph-bib_query?bibcode=2006A\%26A...458%
..477S&db_key=AST}{{Smits}, M., {Maccarone}, T.~J., {Kundu}, A., \& {Zepf},
  S.~E.} 2006, \aap, 458, 477

\bibitem[{{Spitler} {et~al.}(2006){Spitler}, {Larsen}, {Strader}, {Brodie},
  {Forbes}, \& {Beasley}}]{spitler06a}
\href{http://adsabs.harvard.edu/abs/2006AJ....132.1593S}{{Spitler}, L.~R.,
  {Larsen}, S.~S., {Strader}, J., {Brodie}, J.~P., {Forbes}, D.~A., \&
  {Beasley}, M.~A.} 2006, \aj, 132, 1593

\bibitem[{{Strader} {et~al.}(2006){Strader}, {Brodie}, {Spitler}, \&
  {Beasley}}]{strader06a}
\href{http://adsabs.harvard.edu/cgi-bin/nph-bib_query?bibcode=2006AJ....132.23%
33S&db_key=AST}{{Strader}, J., {Brodie}, J.~P., {Spitler}, L., \& {Beasley},
  M.~A.} 2006, \aj, 132, 2333

\bibitem[{{Thomas} {et~al.}(2003){Thomas}, {Maraston}, \& {Bender}}]{thomas03a}
\href{http://adsabs.harvard.edu/cgi-bin/nph-bib_query?bibcode=2003MNRAS.339..8%
97T&db_key=AST}{{Thomas}, D., {Maraston}, C., \& {Bender}, R.} 2003, \mnras,
  339, 897

\bibitem[{{Thomas} {et~al.}(2005){Thomas}, {Maraston}, {Bender}, \& {de
  Oliveira}}]{thomas05a}
\href{http://adsabs.harvard.edu/cgi-bin/nph-bib_query?bibcode=2005ApJ...621..6%
73T&db_key=AST}{{Thomas}, D., {Maraston}, C., {Bender}, R., \& {de Oliveira},
  C.~M.} 2005, \apj, 621, 673

\bibitem[{{Tonry} {et~al.}(2001){Tonry}, {Dressler}, {Blakeslee}, {Ajhar},
  {Fletcher}, {Luppino}, {Metzger}, \& {Moore}}]{tonry01}
\href{http://adsabs.harvard.edu/cgi-bin/nph-bib_query?bibcode=2001ApJ...546..6%
81T&db_key=AST}{{Tonry}, J.~L., {Dressler}, A., {Blakeslee}, J.~P., {Ajhar},
  E.~A., {Fletcher}, A.~., {Luppino}, G.~A., {Metzger}, M.~R., \& {Moore},
  C.~B.} 2001, \apj, 546, 681

\bibitem[{{Tremaine} {et~al.}(2002){Tremaine}, {Gebhardt}, {Bender}, {Bower},
  {Dressler}, {Faber}, {Filippenko}, {Green}, {Grillmair}, {Ho}, {Kormendy},
  {Lauer}, {Magorrian}, {Pinkney}, \& {Richstone}}]{tremaine02a}
\href{http://adsabs.harvard.edu/cgi-bin/nph-bib_query?bibcode=2002ApJ...574..7%
40T&db_key=AST}{{Tremaine}, S., {et~al.}} 2002, \apj, 574, 740

\bibitem[{{van Dokkum}(2001)}]{vandokkum01a}
\href{http://adsabs.harvard.edu/cgi-bin/nph-bib_query?bibcode=2001PASP..113.14%
20V&db_key=AST}{{van Dokkum}, P.~G.} 2001, \pasp, 113, 1420

\bibitem[{{Vesperini} \& {Zepf}(2003)}]{vesperini03a}
\href{http://adsabs.harvard.edu/cgi-bin/nph-bib_query?bibcode=2003ApJ...587L..%
97V&db_key=AST}{{Vesperini}, E. \& {Zepf}, S.~E.} 2003, \apjl, 587, L97

\bibitem[{{Wang}(2004)}]{wang04a}
\href{http://adsabs.harvard.edu/cgi-bin/nph-bib_query?bibcode=2004ApJ...612..1%
59W&db_key=AST}{{Wang}, Q.~D.} 2004, \apj, 612, 159

\bibitem[{{Yoon} {et~al.}(2006){Yoon}, {Yi}, \& {Lee}}]{yoon06a}
\href{http://adsabs.harvard.edu/abs/2006Sci...311.1129Y}{{Yoon}, S.-J., {Yi},
  S.~K., \& {Lee}, Y.-W.} 2006, Science, 311, 1129

\bibitem[{{Zepf} \& {Ashman}(1993)}]{zepf93a}
\href{http://adsabs.harvard.edu/cgi-bin/nph-bib_query?bibcode=1993MNRAS.264..6%
11Z&db_key=AST}{{Zepf}, S.~E. \& {Ashman}, K.~M.} 1993, \mnras, 264, 611

\end{thebibliography}

\end{document}